\documentclass{emulateapj}
\usepackage{natbib}

\begin{document}

\title{Star Formation Rates and Stellar Masses of H$\alpha$ selected Star-Forming Galaxies at z=0.84: A Quantification of the Downsizing.}

\author{V\'{i}ctor Villar, Jes\'us Gallego}
\affil{Departamento de Astrof\'{\i}sica, Facultad de CC. F\'{\i}sicas, Universidad Complutense de Madrid, E-28040 Madrid, Spain}
\email{vvillar,j.gallego@fis.ucm.es}

\author{Pablo G. P\'erez-Gonz\'alez}
\affil{Departamento de Astrof\'{\i}sica, Facultad de CC. F\'{\i}sicas, Universidad Complutense de Madrid, E-28040 Madrid, Spain
Associate Astronomer at Steward Observatory, The University of
Arizona, 933 N Cherry Avenue, Tucson, AZ 85721, USA}
\email{pgperez@fis.ucm.es}

\author{Guillermo Barro, Jaime Zamorano}
\affil{Departamento de Astrof\'{\i}sica, Facultad de CC. F\'{\i}sicas, Universidad Complutense de Madrid, E-28040 Madrid, Spain}
\email{gbarro,jzamorano@fis.ucm.es}

\author{Kai Noeske}
\affil{Space Telescope Science Institute, 3700 San Martin Drive, Baltimore, MD 21218, USA}
\email{noeske@stsci.edu}

\author{David C. Koo} 
\affil{Lick Observatory, University of California, Santa Cruz, CA 95064, USA}
\email{koo@ucolick.org}

\begin{abstract}

In this work we analyze the physical properties of a sample of 153 star forming galaxies at z$\sim$0.84, selected by their H$\alpha$ flux with a NB filter. B-band luminosities of the objects are higher than those of local star forming galaxies. Most of the galaxies are located in the blue cloud, though some objects are detected in the green valley and in the red sequence. After the extinction correction is applied virtually all these red galaxies move to the blue sequence, unveiling their dusty nature. 
A check on the extinction law reveals that the typical extinction law for local starbursts is well suited for our sample but with E(B-V)$_{stars}$=0.55 E(B-V)$_{gas}$. We compare star formation rates (SFR) measured with different tracers (H$\alpha$, FUV and IR) finding that they agree within a factor of three after extinction correction. We find a correlation between the ratios SFR$_{FUV}$/SFR$_{H\alpha}$, SFR$_{IR}$/SFR$_{H\alpha}$ and the EW(H$\alpha$) (i.e. weighted age) which accounts for part of the scatter. We obtain stellar mass estimations fitting templates to multi-wavelength photometry. The typical stellar mass of a galaxy within our sample is $\sim$10$^{10}$M$_{\odot}$. The SFR is correlated with stellar mass and the specific star formation rate (sSFR) decreases with it, indicating that massive galaxies are less affected by star formation processes than less massive ones. This result is consistent with the {\em downsizing} scenario. To quantify this {\em downsizing} we estimated the {\em quenching} mass M$_{Q}$ for our sample at z$\sim$0.84, finding that it declines from M$_{Q}\sim$10$^{12}$M$_{\odot}$ at z$\sim$0.84 to M$_{Q}\sim$8$\times$10$^{10}$ M$_{\odot}$ at the local Universe.

\end{abstract}


\section{Introduction}

During the last two decades the cosmic Star Formation History of the Universe has been widely studied in order to better constraint galaxy formation and evolution models. Large-area surveys as well as the use of larger telescopes have consolidated our knowledge at low-intermediate redshifts (z=0.0-1.0) \citep[see][]{Hop06}. Several measurements exist at higher redshifts \citep{PG05,Geach08, Reddy09, Hayes10} and at present we have started to probe the most distant Universe at z$\sim$7--8 \citep{Bouwens09,Bouwens10arXiv}.

In general, the Star Formation Rate density (SFRd) history from the Local Universe to z$\sim$1 is well accepted. Near z$\sim$1, where the rise in SFRd from the local Universe slows down, there are several measurements obtained through samples selected in a variety of ways and using different Star Formation Rate (SFR) tracers. Results from H$\alpha$, UV and IR agree reasonably, although with higher dispersion than at lower redshifts \citep{Garn10} . This scattering is originated (at least partially)  due to the two aforementioned processes: the sample selection or/and the SFR estimation. The estimations of the SFRd at this redshift have been measured mainly through UV \citep{Lilly96,Connolly97,Cowie99,Wilson02,Schimi05}, IR \citep{Flores99,Lefloch05,PG05} and H$\alpha$ \citep{gla99,Yan99,Hop00,Tresse02,Do06,Villar08,Sobral09, Ly11}. Therefore, it is important to constraint the potential differences that arise when one or another tracer is used to estimate SFRs. 

Estimations of SFR through the UV flux are very sensitive to the extinction correction. The UV also probe older populations than H$\alpha$, thus being more sensitive to recent star formation history \citep{Cal05}. The IR on the other hand is not affected by extinction but is very model dependent \citep{Barro11,Barro11b}. Moreover, it is not well understood how the old population of stars contributes to the IR emission, but it could be a significant fraction \citep{DaCunha08,Salim09}. 
The H$\alpha$ line is one of the best estimators as it is sensitive only to very young stars, not affected by recent star formation history, which may still be detected by UV or IR. It has the problem that at z$>$0.5 it moves to the near Infrared (nIR) domain, where large amounts of spectroscopy are still difficult to obtain, though several nIR multi-object spectrographs for 8-10 meter class telescopes are coming in the next years. In addition, different methods have been used for measuring the total H$\alpha$ line flux, which could lead to discrepancies if some effects are not properly corrected. On one hand we have spectroscopy, long- or multi- slit or through fibers, where aperture corrections are needed to recover the total flux \citep[see for example][]{Do06,Erb06}. On the other hand we have the slitless spectroscopy \citep{Yan99,Hop00} and narrow band \citep{Villar08,Sobral09,Ly11} techniques, which have the advantage that the total flux of the object is recovered and no aperture corrections are needed. Although it is needed to correct from the Nitrogen contribution if the filter is not narrow enough, this effect is easier to estimate than aperture corrections. Obviously, the extinction is also important at this wavelength, though less than in the UV.

Thus, the H$\alpha$ estimator is the best suited to study the instantaneous star formation. As mentioned before the problem is that at this redshift the line is observed in the nIR and little data is available up to date. Then, it is also interesting to asses if UV and IR provide SFRs comparable to those of H$\alpha$, given the large amount of data in these wavelengths available today. 

The selection methods of the samples are also different and target different populations. Selecting the sample in the UV for example, implies a bias against very obscured galaxies, which may not be detected unless very deep observations are carried out. On the other hand, the IR selected samples will favor the selection of objects with large amounts of dust, missing the blue and dust-free objects. Selections based in the H$\alpha$ line will select the objects with ongoing star formation and thus only star forming galaxies (and AGN) will be selected. 

Thus, to study the population of galaxies that are actively forming stars it is necessary to have a well defined sample of star forming galaxies and one of the best technique today is the use of narrow band filters targeting H$\alpha$. 

A population of star forming galaxies selected in this way is ideal to study the SFR sequence, which refers to the correlation that exists between SFR and stellar mass. This correlation has been found at a wide range of redshifts although there has been found evolution with respect to this parameter \citep[see][for a review]{Dutton10}. While the slope is almost constant, the SFR zeropoint increase from the local Universe to redshift z$\sim$2. From this redshift on, the trend remains almost constant with little evolution. There exist some discrepancy in the slope, which is somewhat lower than unity in all cases \citep[see for example][]{Noeske07a,Elbaz07,Salim07}. 

However, some works do not find this correlation. \cite{Caputi06} do not find any trend for their MIPS selected sample at z$\sim$2. More recently, \cite{Sobral10} did not find any evidence of this sequence for their HiZELS sample at z$\sim$0.84, selected with a narrow-band filter targeting H$\alpha$. 

This correlation implies that the slope of specific SFR (sSFR) versus mass is higher than -1. Obviously, if no correlation between stellar mass and SFR is found the slope of sSFR vs. stellar mass is simply -1. The slope of this correlation is very important, as it tells us how decreases the importance of star formation over the already formed stellar mass.

In this work we use a narrow band selected sample of star forming galaxies at z$\sim$0.84, presented in \cite{Villar08}, to compare SFRs obtained from different tracers and to study the relation between stellar mass and SFR. The sample is very well suited for this study as: i) it is directly selected by star formation, so we are not biasing the population of star forming galaxies; ii) the use of the narrow band filter technique provides us reliable H$\alpha$ SFRs to compare with FUV and IR estimations.

This paper is structured as follows. In Section~\ref{sec:sample}, we present the sample and the available datasets. In Section~\ref{sec:agn}, we describe the methods used to get rid of AGN contaminants. Absolute magnitudes and color are presented in Section~\ref{sec:photo}. Section ~\ref{sec:sfr} presents a comparison of SFRs obtained through different estimators as well as a check on the extinction law more suited to our sample. Section~\ref{sec:mass} presents the stellar masses and their relation with star formation rate. Finally, we summarize our results and conclusions in Section~\ref{sec:conclusions}.

Throughout this paper we use AB magnitudes. We adopt the cosmology {\em H$_{0}$} = 70 km s$^{-1}$ Mpc$^{−1}$, $\Omega _{m}$ = 0.3 and $\Omega _{\lambda}$ = 0.7.

\section{Data}
\label{sec:sample}

\subsection{Sample}

This paper analyzes an H$\alpha$ selected sample of galaxies at z=0.84. The
objects are selected by their emission in the H$\alpha$+[N{\sc II}] line and are thus
selected due to intense star formation (except when activity at nuclei level is
present). The sample was first described in \citet[ hereafter V08]{Villar08} \defcitealias{Villar08}{V08} and the reader is referred to
that paper for full details on the sample selection criteria. A brief summary of the process is presented here.

The sample was built using narrow and broad band images in the J band of the near infrared. The narrow-band filter used in this work is J-continuum (Jc) centered at 1.20$\mu$m, corresponding to H$\alpha$ at z=0.84. The search was performed using the
near-infrared camera OMEGA-2000\footnote{http://www.mpia-hd.mpg.de/IRCAM/O2000/index.html}
of the 3.5m telescope at the Calar Alto Spanish-German Astronomical Center (CAHA). OMEGA-2000 is equipped with a
2k$\times$2k Hawaii-2 detector with 18$\mu$m pixels (0$\arcsec$45 on
the sky, 15$\arcmin \times$15$\arcmin$ field of view). Three pointings were
observed, two in the Extended Groth Strip (EGS) and another one in the GOODS-North field, covering a whole area of $\sim$0.174 deg$^2$, and reaching 70\% completeness at a line flux of $\sim$1.5$\times$10$^{-16}$ erg s$^{-1}$ cm$^{-2}$. 

In a first step, 239 emission line candidates were selected (once excluded the stars) by their flux excess in the narrow band, showing a J-J$_{C}$ colour excess significance n$_{\sigma}$ $>$ 2.5 in one or several apertures. Spectroscopic and photometric redshifts were then used to rule out contaminants, either emission line galaxies at other redshifts or objects selected by spectral features or noise. First, the sample was cross-checked against redshift catalogs on GOODS-N and EGS fields. A total of 76 objects were confirmed as genuine H$\alpha$ emitters in the narrow band redshift range, 43 in the Extended Groth Strip and 33 in GOODS-N. Contaminants were mainly emission line galaxies at other redshifts, including a small sample of [O{\sc III}]$\lambda\lambda$4959,5007 emitters at z$\sim$1.4, and a few objects not selected by line emission. The accuracy selecting emission line galaxies was very high, around $\sim$90\%.  Spectroscopic redshifts were only available for 98 objects in the sample, therefore photometric redshifts were used to get rid of interlopers for the rest of the sample. The quality of the estimated photometric redshifts was very high, with 86\% of the objects in the whole EGS and 90\% of the objects in GOODS-N (with reliable spectroscopic redshifts) within $\sigma_{z}/(1+z)<$0.1. Considering the photometric redshifts, a total of 89 objects, 64 in the EGS and 25 in GOODS-N, were added to the final sample. 

Since the original paper was published new spectroscopic data has been made public, increasing the number of confirmed sources in 18 objects, for a total of 94. Only two objects are found to be incorrectly classified as H$\alpha$ emitters, which has been removed from the sample. Thus, the selection efficiency found in the original sample remains very similar.

The final sample of H$\alpha$ emitters at z=0.84 contains 165 objects, 107 in the EGS and 58 in GOODS-N, 94 (57 \%) of them confirmed by optical spectroscopy (after including three objects with low quality spectroscopic redshift).  However, due to insufficient complementary data, we have discarded 12 objects. Hence, the sample used in this paper is composed of 153 objects.

Line fluxes have been recomputed using the formalism described in \cite{P07} (hereafter P07), introducing the Nitrogen contribution in the filters effective widths. This forces us to assume an initial value for the nitrogen contribution, setting it to the average value found in \citetalias{Villar08}: $I([NII]\lambda6584)/I(H\alpha)$=0.26. This provides an initial estimate of the H$\alpha$ line flux, without nitrogen contribution. With the equivalent width, we can estimate the nitrogen contribution, given the correlation between EW(H$\alpha$) and $I([NII]\lambda6584)/I(H\alpha)$ found in the local SDSS sample (see P07). We then reestimate the nitrogen contribution and compute again the line flux and equivalent width. The latter provides a new estimation of the Nitrogen contribution. The process is repeated until it converges, usually in two or three steps.

The line flux estimation through the narrow and broad band filters assumes a most likely redshift for the object, i.e. we assume a redshifted wavelength for the emission line based in the shape of the filter and in the cosmology \citep[see][]{P07}. In fact, the objects distribute along the wavelength range covered by the narrow band filter, being most likely detected near the filter's central wavelength, where the transmission is high. Thus, assuming a most likely redshift for the objects seems reasonably for the majority of them. However, for the objects that are selected in the wings of the filter, where the transmission falls abruptly, the recovered fluxes differ significantly. Even in the regions of high transmission, there could be important effects due to the presence of the nitrogen lines. We can correct these effects introducing the real redshift, which we know for half the sample, in the equations to compute the line flux \citep[see][]{P07}. 

\begin{figure}[t]
\includegraphics[width=0.5\textwidth]{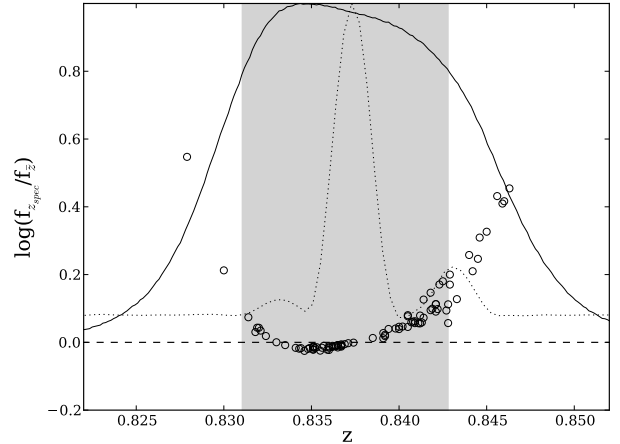}
\caption{\label{FLUX_COMP} Line flux ratio between those computed with the individual spectroscopic redshift and with the average redshift. The narrow band filter shape is represented by the continuous line. The shaded region comprises the redshift range where the filter transmission is above 80\%, where most of the objects are detected. The dotted line represents the H$\alpha$ line surrounded by the Nitrogen lines.}
\end{figure}

In figure~\ref{FLUX_COMP} we compare the line fluxes estimated with the real redshift versus the ones estimated in the general way. The narrow-band filter shape is also shown as reference. It can be seen that, for the objects that fall in the wings of the filter, the line flux is clearly subestimated. It is worth noting also that a significant line flux fraction is lost even when the transmission is high, as is the case of the objects with higher redshifts within the shaded region in the figure, where the transmission is always above 80\%. This is due to the fact that the [N{\sc II}]$\lambda$6584\AA\ line, which is the most intense of the two nitrogen lines, shifts to wavelengths where the transmission of the filter is very reduced, and hence, if we do not take this effect into account, we overcorrect the nitrogen contribution, estimating fainter line fluxes. The effect is also present at shorter wavelengths, though in that case is the other Nitrogen line ([N{\sc II}]$\lambda$6548\AA) the one shifted to wavelengths with lower transmission. However, as this line is 3$\times$ weaker than the other one, the effect is less pronounced. 

The amount of extinction for each galaxy was estimated through the FIR to UV flux ratio or the UV slope when the FIR data was not available (see V08). We used the extinction law derived by \cite{Cal00}. As new data is available now, specially regarding MIPS 24$\mu$m, we have recomputed the extinctions for all the objects, considering, in addition, the results on the check on the extinction law (see section~\ref{subsec:dust}). The median extinction for our sample is 1.24$^{m}$ in H$\alpha$, adopting values between 0$^{m}$ and 3.8$^{m}$. Once applied these corrections, H$\alpha$ luminosities and star formation rates were computed.

\subsection{Additional data}

In order to estimate the different properties analyzed in this paper, we use several additional data sets sampling a wide range of the electromagnetic spectrum, from the Far Ultraviolet (GALEX FUV) to the Mid Infrared (MIPS 24$\mu$m). These complementary data sets have been collected as part of the UCM Rainbow database \citep[see V08,][ for details]{PG08,Barro11}, and have been gathered in part by the AEGIS \citep{davis07} and GOODS \citep{Dick01} projects.

Briefly, in the EGS we have used optical data gathered with {\em MegaCam} at the 4m Canada France Hawaii Telescope (CFHT), covering the following bands: {\em u$^{*}$}, {\em g'}, {\em r'}, {\em i'}, and {\em z'}. We have also used B, R, and I images obtained with the same telescope but with the camera CFHT 12k, as described in \cite{Coil04}. Deep optical R band data taken with SuprimeCam at Subaru 8m as part of the Subaru Suprime-Cam Weak-Lensing Survey \citep{Miya07} are also available. In the domain of the near infrared, images in the J band were obtained with Omega2000, as part of the data necessary to make the sample selection, and K-band images obtained with Omega prime \citep{Barro09}. Both instruments were located at the 3.5m telescope at CAHA. Space-based optical images acquired with the Advanced Camera for Surveys (ACS) on board {\em HST} are available in two bands: {\em F606W} and {\em F814W} (hereafter {\em V$_{606}$} and {\em i$_{814}$}). The {\em Galaxy Evolution Explorer} \citep[GALEX, ][]{Martin05} provides ultraviolet deep images in the far ultraviolet (FUV; 153 nm) and the near ultraviolet (NUV; 231 nm). The space observatory {\em Spitzer} observed the EGS field at 3.6, 4.5, 5.8, and 8 $\mu$m with the IRAC instrument and in 24 $\mu$m with MIPS \citep{Barmby08}.

In GOODS-N we made use of deep optical and near infrared images \citep[UBVRIzHK$_{s}$, ][]{capak04}, as well as our own J- and K-band images, both of them obtained with Omega2000 \citep[see V08 and][for details]{Barro09}. As in the case of the EGS, space based observatories provide us with ultraviolet and infrared data, as well as high resolution additional optical data. GALEX observed the region in Far- and Near- Ultraviolet channels. {\em Spitzer} observed in the mid infrared (3.6 to 8 $\mu$m; IRAC) and in the far infrared (24 $\mu$m; MIPS). The ACS on board the HST adds optical data in four bands {\em F435B} ({\em B$_{435}$}), {\em F606W} ({\em V$_{606}$}), {\em F775W} ({\em i$_{775}$}), and {\em F850LP} ({\em z$_{850}$}).

\section{AGN Contaminants}
\label{sec:agn}

The selection of a sample through the H$\alpha$ line is sensible to Active Galactic Nuclei (AGN) contamination, as they are also powerful emitters in this line. Although both AGN and SFR could contribute together to the flux, disentangling both components is out of the scope with the available data. Thus, we will remove the objects classified as AGN from our sample.

In this work we detect the presence of 13 (8\%) AGN using two complementary methods: X-ray luminosity and mid-IR colors. 

\subsection{X-ray luminosity}
We have cross-correlated our sample with the available X-ray catalogs in the EGS and GOODS-N fields. In the EGS fields we have the {\em AEGIS-X} X-ray catalog \citep{Laird09}, which covers a large area within the EGS. Observations were made with the {\em Chandra} X-ray observatory with nominal exposure time of 200ks. In GOODS-N we have used the catalog created by Laird et al. from observations taken by {\em Chandra} with an exposure time of 2Ms \citep{Alexander03}. We find three X-ray counterparts in the EGS and four in GOODS-N, within a 2$\arcsec$ search radius. The three objects in the EGS present high X-ray fluxes (L$_{X}>$6$\times$10$^{42}$erg\ s$^{-1}$), revealing their AGN nature. In GOODS-N, due to the depth of the observations, we find three objects whose X-ray luminosities are compatible with an star formation origin. The derived SFRs, using the calibration given by \cite{Ranalli03}, agree within a factor of three with the H$\alpha$ derived ones. Thus we have only discarded the four objects with X-ray luminosities whose origin could only be attributed to an AGN.

\subsection{Mid Infrared colors}

\begin{figure}[t]
\includegraphics[width=0.5\textwidth]{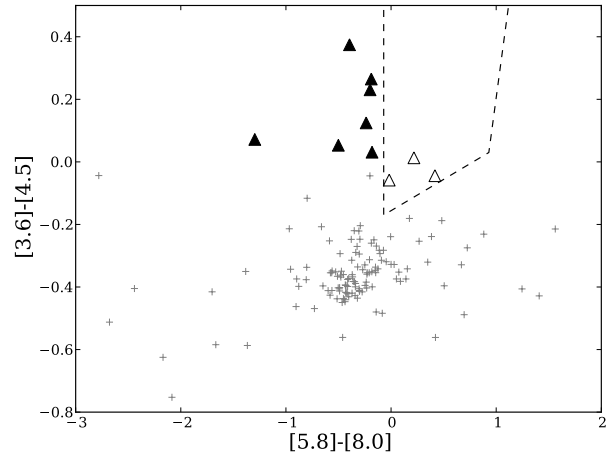}
\caption{\label{AGN_MIDIR} {\em IRAC} color-color plots for the selected sample with photometry in the four bands. The wedge delimited by the dashed polygon encloses the emitters powered by an AGN (open triangles). Seven objects (filled triangles), that fall outside this wedge, with positive [3.6]-[4.5] color have been also considered AGNs because they have a rising SED and the errors make them compatible with being located inside the wedge. }
\end{figure}

Some AGN are heavily obscured and even the deepest X-ray observations can miss a significant fraction of them \citep{Park10}. In this case the X-ray emission is absorbed by the circumnuclear dust and re-emitted in the infrared. One way to detect these obscured AGNs is looking at their mid-IR colors.    
The color criterion defined by \cite{Stern05} is based in the differences in the mid IR emission shown by star-forming galaxies and AGNs. The star-forming galaxies SED peaks at 1.6$\mu$m, falling at longer wavelengths \citep{Garn10}. In the case of an AGN the emission do not decreases at longer wavelengths, due to the re-emission of light absorbed by the circumnuclear region in the mid-IR. Unfortunately the distinction becomes less pronounced at our redshift, as pointed out by \cite{Stern05}.

In figure~\ref{AGN_MIDIR} we show the objects that fulfill the criterion (inside the dashed polygon), which are AGNs, as well as the rest of the objects, which are pure star forming galaxies. A total of three galaxies fall within the wedge defined by \citeauthor{Stern05} Seven other objects with positive [3.6]-[4.5] color fall relatively close to this wedge (except one), although do not fulfill the criterion. We have decided to consider these objects as AGN contaminants, given that photometry errors could have placed them outside the region and that they have a rising SED ([3.6]-[4.5]$>$0). Only one of these ten objects have an X-ray counterpart.

We note that this classification could be selecting star forming galaxies instead of AGNs, as pointed out by \cite{Donley08}. We decided to exclude them all to make sure we are not introducing any AGN. 

Another way to check the presence of obscured AGN is through the power-law criterion \citep{AlonsoH06}. We do not find any galaxy showing this characteristic power-law shape in the mid-IR SED.

\section{Photometric properties}
\label{sec:photo}

\begin{figure}[t]
\includegraphics[width=0.5\textwidth]{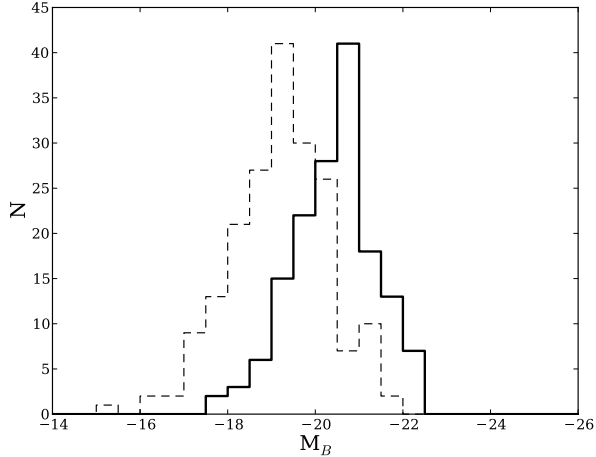}
\caption{\label{B_HIST} Histogram of rest-frame B-band absolute magnitudes. The thick line shows the distribution for the z$\sim$0.84 sample. The dashed line correspond to the UCM local sample, selected also by their H$\alpha$ emission.}
\end{figure}

The histogram of rest-frame B-band magnitudes for our sample is shown in figure~\ref{B_HIST}. The median of the distribution is M$_{B}$=-20.5$^{m}$, reaching the most luminous objects M$_{B}$=-22.5$^{m}$. The standard deviation of the distribution is 0.9$^{m}$. For comparison we also show in the figure the UCM local sample of star forming galaxies \cite{Zam94,Zam96}, selected also by the H$\alpha$ line flux. In the local sample no galaxies brighter than M$_{B}\sim$-22 are detected, suggesting that star forming galaxies at z$\sim$0.84 are more luminous than their local analogous. The volume sample in each survey is very different: 10$^{5}$ Mpc$^{3}$ for the typical object in the UCM, while for our sample the surveyed volume is $\sim$15$\times$10$^{3}$ Mpc$^{3}$. However, given the larger volume explored in the UCM survey with respect to our survey, it is clear that star forming galaxies at z$\sim$0.84 are in general brighter in the B band.

\begin{figure}[t]
\includegraphics[width=0.5\textwidth]{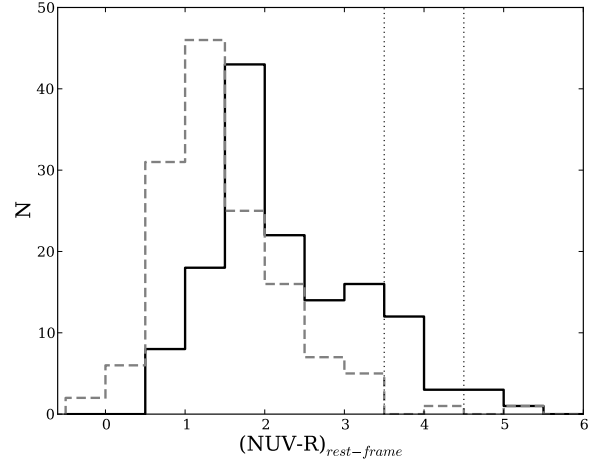}
\caption{\label{NUVR_HIST} Histogram of rest-frame NUV-R colors for our sample. The black line represents the sample without applying the correction for extinction. The dashed line shows the distribution once the extinction has been corrected. It can be seen that most objects fall in the blue cloud ((NUV-R)$<$3.5). Before the extinction correction is applied some objects fall in the green valley (3.5$<$ (NUV-R)$<$4.5) and in the red sequence ((NUV-R)$>$4.5). After it is applied only two objects remain outside the blue cloud.}
\end{figure}

A clear bimodality in the color of galaxies was first found by \cite{Strateva01} analyzing the optical colors of the SDSS sample. Galaxies divide mainly in two groups: the blue cloud and the red sequence. The blue cloud is populated by star forming galaxies whereas galaxies with no recent star formation fill the red sequence. This bimodality is also present at higher redshifts \citep{Willmer06,Faber07}. An intermediate region, the green valley, was identified using the NUV-R color \citep{Wyder07}. The galaxies within this group are either in a transition phase from the blue cloud to the red sequence, due to the shutdown of star formation; or are star forming galaxies with high extinction \citep{Martin07,Salim07}. In figure~\ref{NUVR_HIST} we depict the NUV-R color for our sample. Most of the sample belong to the blue cloud (NUV-R$<$3.5), in agreement with their star-forming nature. However, some galaxies fall in the green-valley (3.5$<$NUV-R$<$4.5) and a few of them in the red sequence (NUV-R$>$4.5). This can be explained due to the high extinction present in these galaxies and, indeed, when the extinction is corrected only two galaxies fall outside the blue cloud. One of them is not confirmed by optical spectroscopy so it might not be a real z$\sim$0.84 emitter. In fact, its photo-z $\chi^{2}$ distribution does not present a clear peak at that redshift but a flatter distribution. The other one, although confirmed by optical spectroscopy, is very close to another galaxy ($<$2$\arcsec$) and its photometry might be affected.

\section{Star Formation Rates}
\label{sec:sfr}

This work uses the H$\alpha$ luminosity as the principal estimator of the instantaneous SFR of galaxies. The H$\alpha$ line flux has been used to select the sample, making it very suitable to study star formation processes, as it has been selected by this property. However, star formation involves physical processes whose imprint become observable along a wide range of the electromagnetic spectrum: X-rays, ultraviolet, forbidden recombination lines ([O {\sc II}]$\lambda$3727), far infrared, radio, etc. In this work, in addition to H$\alpha$, given the depth and coverage of the multi wavelength data available, we estimate SFRs through Far Ultraviolet and Far Infrared luminosities. Each tracer is affected by different phenomenons and is originated by different physical mechanisms, related (at least in part) with star formation processes. Thus, the different results obtained with different tracers could yield some information about the properties of the galaxy that hosts the star formation processes.

The H$\alpha$ line is produced due to the recombination processes in ionized hydrogen present in the clouds of gas and dust that surround the newly formed stars. The massive stars of type O and B produce an intense radiation field capable of ionize the hydrogen atoms. When the equilibrium is reached, the recombination of the free electrons with the ionized hydrogen produce several emission lines, being the H$\alpha$ line one of the most luminous in the visible. To obtain the star formation rate from the H$\alpha$ luminosity (L$_{H_{\alpha}}$) we apply the relation given by \cite{Ken98}:

\begin{equation}
\label{SFR_Halpha:eq}
SFR_{H\alpha}(M_\odot \, yr^{-1})=7.9\times10^{-42 }L_{H_{\alpha}}(erg\, s^{-1})
\end{equation} 
 
\noindent where a \cite{Salpeter55} IMF has been considered. H$\alpha$ traces directly the SFR and has very low dependence on metallicity or on ionization conditions of the gas cloud. Among the adverse effects the most important are the extinction and the escape fraction of ionizing photons. The former is common to optical indicators and extinctions as high as $\sim$4 magnitudes in the H$\alpha$ line can be found in our sample, although the median extinction is 1.24 magnitudes. The latter implies a subestimation of the star formation rate if the escape fraction is high. Fractions up to 50\%\ have been measured for individual HII regions \citep{Oey97}. However, this fraction turns out to be much lower when the whole galaxy is considered (which is our case) decreasing to less than 3\%\ as measured by \cite{Lei95}. Another adverse effect recently shown by \cite{Lee09} is the underestimation of star formation rate for dwarf galaxies. Nevertheless, this effect appears for star formation rates below 0.03 M$_{\odot}$yr$^{-1}$, two orders of magnitude lower than our lowest star formation rate, so it does not affect the estimations for our samples.

Ultraviolet emission comes directly from young massive stars formed in the star formation region. To compute the UV SFRs we use the following calibration \cite{Ken98}:

\begin{equation}
\label{SFR_UV:eq}
SFR_{FUV}(M_{\odot}\, yr^{-1})=1.4\times10^{-28}L_{FUV}(erg\; s^{-1} Hz^{-1})
\end{equation} 

\noindent where $L_{FUV}$ is the FUV luminosity spectral density. Although we apply it to the UV flux in 1500\AA, the calibration is valid in the 1500-2800 \AA\ range, as the spectrum is nearly flat in that regime. 

The dust in a galaxy absorbs part of the radiation emitted at short wavelengths and re-emits it in the IR. This absorption is more intense at shorter wavelengths. Given that young stars radiate most of their luminosity in the ultraviolet, there exists a correlation between IR luminosity and star formation. The correlation between luminous regions in H$\alpha$ and in IR confirmed the validity of the latter as a valid star formation tracer \citep[see ][ for details]{Dever97}. More recently, observations carried out with the Spitzer space telescope, with improved resolution, confirmed those results \citep{Cal05,PG06}. 

The calibration of the infrared emission as a star formation tracer is not simple as it depends on several factors: geometrical distribution and optical thickness of the dust, fraction of emission coming from old stars, etc. In the ideal case the star forming regions would be surrounded by dust clouds, opaque enough to re-radiate all the region luminosity. However, these ideal conditions differ from the real scenario, due in part to the aforementioned factors. This complicates the calibration and increases the dispersion.

Given the infrared luminosity L$_{IR}$(8-1000$\mu$m), the following relation can be used to estimate the star formation rate \citep{Ken98}:

\begin{equation}
\label{SFR_IR:eq}
SFR_{IR}(M_{\odot} yr^{-1})=1.71\times10^{10}L_{IR}(L_{\odot}) 
\end{equation}
 
\noindent where L$_{IR}$ corresponds to the total infrared luminosity between 8 and 1000 $\mu$m.

The great advantage of this tracer is that it is not affected by extinction. However, the estimations are sensitive to other factors like the dust spatial distribution, old stellar population contribution, etc. Moreover, it is necessary to estimate the total infrared luminosity in the range 8-1000 $\mu$m. In general, this is done with a few measurements in the infrared, usually with wavelengths shorter than 24$\mu$m. Recent works have demonstrated that a better relation exists between H$\alpha$ and flux at 24$\mu$m \citep[see, for example, ][]{PG06,Cal07,Rieke09, Ken09}. However, this wavelength is not available for our z=0.84 sample, because the MIPS observed 24$\mu$m data turns into restframe $\sim$13$\mu$m.

\subsection{Dust attenuation}
\label{subsec:dust}

In order to properly compare the SFRs it is necessary to correct the effect introduced by the extinction. In V08 we computed the extinction using the dust flux to UV flux ration and the UV slope when the IR data was not available. This provided us the attenuation in the FUV band, from which the attenuation at the H$\alpha$ line was estimated assuming a \cite{Cal00} extinction law, given the star-forming nature of our sample. This law considers that the nebular emission is more extincted than the stellar emission: E(B-V)$_{stars}$= $\gamma$ E(B-V)$_{gas}$, with $\gamma$=0.44. However, this factor may be different for different populations and/or dust geometry and may depend on redshift. \cite{Garn10} found for the S09 sample that $\gamma\sim$0.5, slightly higher than the typical value. \cite{Erb06} found that, in order to reconcile their SFR estimations in the UV and H$\alpha$ for their z$\sim$2 UV selected sample, the same color excess has to be affecting both the gas and stars, i.e. $\gamma\sim$1. \cite{Yoshi10}, using a sample of BzK selected galaxies at z$\sim$2, found that their data was consistent with the original $\gamma$=0.44 value, although galaxies with low SFRs are consistent with $\gamma$=1.

The $\gamma$ factor arises from the fact that the UV and the nebular emission have different spatial origins, due to the different population of stars each one is tracing. Whereas the nebular emission is originated from very massive and young stars, the UV emission is originated by less massive and older stars. Thus, this factor might be different depending on the star formation history of the galaxies. Moreover, the extinction law may be different than that of \cite{Cal00}. Thus, it is important to estimate this value for our sample and, if possible, to verify the suitability of the \citeauthor{Cal00} (with the same or different $\gamma$) extinction law for our sample. 

We tackle this problem estimating the extinction law in the UV regime and in the H$\alpha$ line. Thanks to the large amount of optical broad-band data available we can obtain several estimations of the SFR (affected by extinction) at different wavelengths within the UV, in addition to the H$\alpha$ estimation. If we assume that every different SFR estimation, once corrected for extinction, shall give the same SFR, we can obtain the extinction in each wavelength comparing to the total SFR:

\begin{eqnarray}
SFR^{total} &=& SFR^{uncor}_{UV_{n}} 10^{0.4\, \kappa(UV_{n})\, E(B-V)_{stars}} \\
SFR^{total} &=& SFR^{uncor}_{H\alpha} 10^{0.4\, \kappa(H\alpha)\, E(B-V)_{gas}}
\end{eqnarray}

\noindent where as SFR$^{total}$ we use the SFR given by the IR(8-1000$\mu$m), UV$_{n}$ represents each different UV wavelength. Thus,  
we can obtain $\kappa$($\lambda$) for different wavelengths:

\begin{eqnarray}
\kappa(UV_{n}) &=& \frac{2.5}{E(B-V)_{stars}}\, log \left ( \frac{SFR^{total}}{SFR^{uncor}_{UV_{n}}} \right ) \\
&=& \frac{2.5}{\gamma\, E(B-V)_{gas}}\, log \left ( \frac{SFR^{total}}{SFR^{uncor}_{UV_{n}}} \right ) \\
\kappa(H\alpha) &=& \frac{2.5}{E(B-V)_{gas}}\, log \left ( \frac{SFR^{total}}{SFR^{uncor}_{H\alpha}} \right )
\end{eqnarray}

\noindent where we have everything related to the color excess in the gas E(B-V)$_{gas}$ through the $\gamma$ factor. At this point we are interested in the shape of the extinction law and is therefore necessary to get rid of the amount of extinction, parametrized by E(B-V)$_{gas}$ for each galaxy. We normalize then by the value at 6563\AA:

\begin{eqnarray}
\kappa_{6563}(UV_{n}) &=& \frac{1}{\gamma}\, \frac{\displaystyle log \left ( \frac{\displaystyle SFR^{total}}{\displaystyle SFR^{uncor}_{UV_{n}}} \right )}{\displaystyle log \left ( \frac{\displaystyle SFR^{total}}{\displaystyle SFR^{uncor}_{H\alpha}} \right )} \\
\kappa_{6563}(H\alpha) &=& 1
\end{eqnarray}

\noindent where $\gamma$ is not known and therefore we can only measure empirically  $\gamma \, \cdot\, \kappa$(UV$_{n}$). If we fit these values to a extinction law we can obtain this factor, which was our original goal.

In our sample there are data available for four bands in the UV rest-frame within the range 1900-3000\AA, where the spectrum is almost flat once the dust effect has been removed and we can use the \cite{Ken98} calibration. In both EGS and GOODS-N there are observations at $\sim$1950\AA\ and $\sim$2400\AA. There exist an aditional third band in each field but with different wavelength: $\sim$2650\AA\ in EGS and $\sim$2950\AA\ in GOODS-N. Thus, we sample four different wavelengths in the UV. There are 72 objects for which all the UV and infrared needed data is available. 

There are two things we want to check: a) which is the $\gamma$ factor appropriate for our sample? and b) is the \citeauthor{Cal00} extinction law suitable for our sample? To answer these questions we fit two different extinction laws to the data: the aforementioned \citeauthor{Cal00} extinction law and the \cite{Cardelli89} one. The fitting process gives us the $\gamma$ factor needed to make the UV and H$\alpha$ measurements consistent and it allow also to check which extinction law provide a better agreement with the data, through the computation of the $\chi^{2}$ value of each fit.

\begin{figure}[t]
\includegraphics[width=0.5\textwidth]{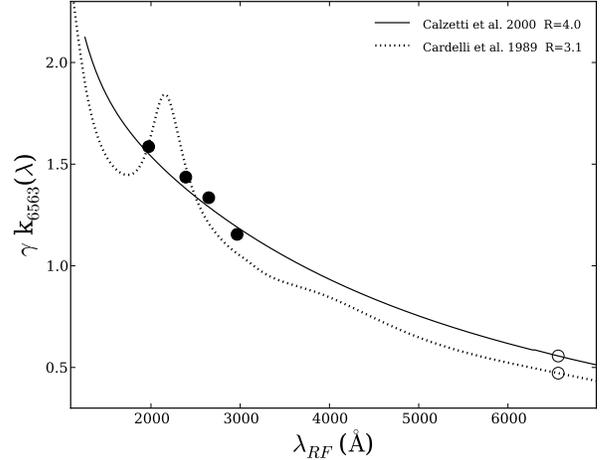}
\caption{\label{EXT_CHECK} Derived reddening curve for our sample (filled circles). The continuos line represents the Calzetti extinction law with $\gamma$=0.55. The dotted line corresponds to the Cardelli extinction law with $\gamma$=0.46. The open circles represent the reddening that would have the stellar continuum for each extinction law and their corresponding $\gamma$ values.   }
\end{figure}

In figure~\ref{EXT_CHECK} we show the $\gamma \, \cdot\, \kappa$(UV$_{n}$) obtained computing the median of the different UV measurements. We also show the Calzetti and Cardelli extinction laws that best fit to the data points. The Calzetti extinction law is more consistent with our measurements. We obtain the following $\chi^{2}$ values for each fit: $\chi^{2}_{Cal00}$=0.2 versus $\chi^{2}_{Car89}$=0.6. These values are below one, due to the large errors we are working with. We consider that the \citeauthor{Cal00} law is best suited for our sample, as the residuals are lower and we are dealing with the same uncertainties. In both cases a heavier attenuation in the nebular gas than in the stellar continuum is needed, i.e a $\gamma$ factor lower than one. We obtain $\gamma_{Cal00}$=0.55$\pm$0.20 and $\gamma_{Car89}$=0.46$\pm$0.17. If we do the analysis on a galaxy by galaxy basis we obtain similar results. The extinction analysis on individual galaxies and the relation with other properties will be presented in a future paper.

If we repeat this process discarding all objects not confirmed by spectroscopy we obtain very similar results. In this case the number of galaxies is reduced to 57 and we obtain $\gamma_{Cal00}$=0.56$\pm$0.20 and $\gamma_{Car89}$=0.47$\pm$0.17. The $\chi^{2}$ values for each fit are now: $\chi^{2}_{Cal00}$=0.4 versus $\chi^{2}_{Car89}$=1.4. A comparison among results (from this and other sections) for the whole sample and that only containing spectroscopically confirmed objects is shown in Table\ref{tab:comp}. 

To summarize, the \citeauthor{Cal00} extinction law is well suited for our sample with $\gamma$=0.55, a value slightly higher than the original 0.44 value. We have assumed this extinction law with this $\gamma$ factor on the dust attenuation estimations.

\subsection{Comparison of SFR tracers}

\subsubsection{Ultraviolet}

\begin{figure}[t]
\includegraphics[width=0.5\textwidth]{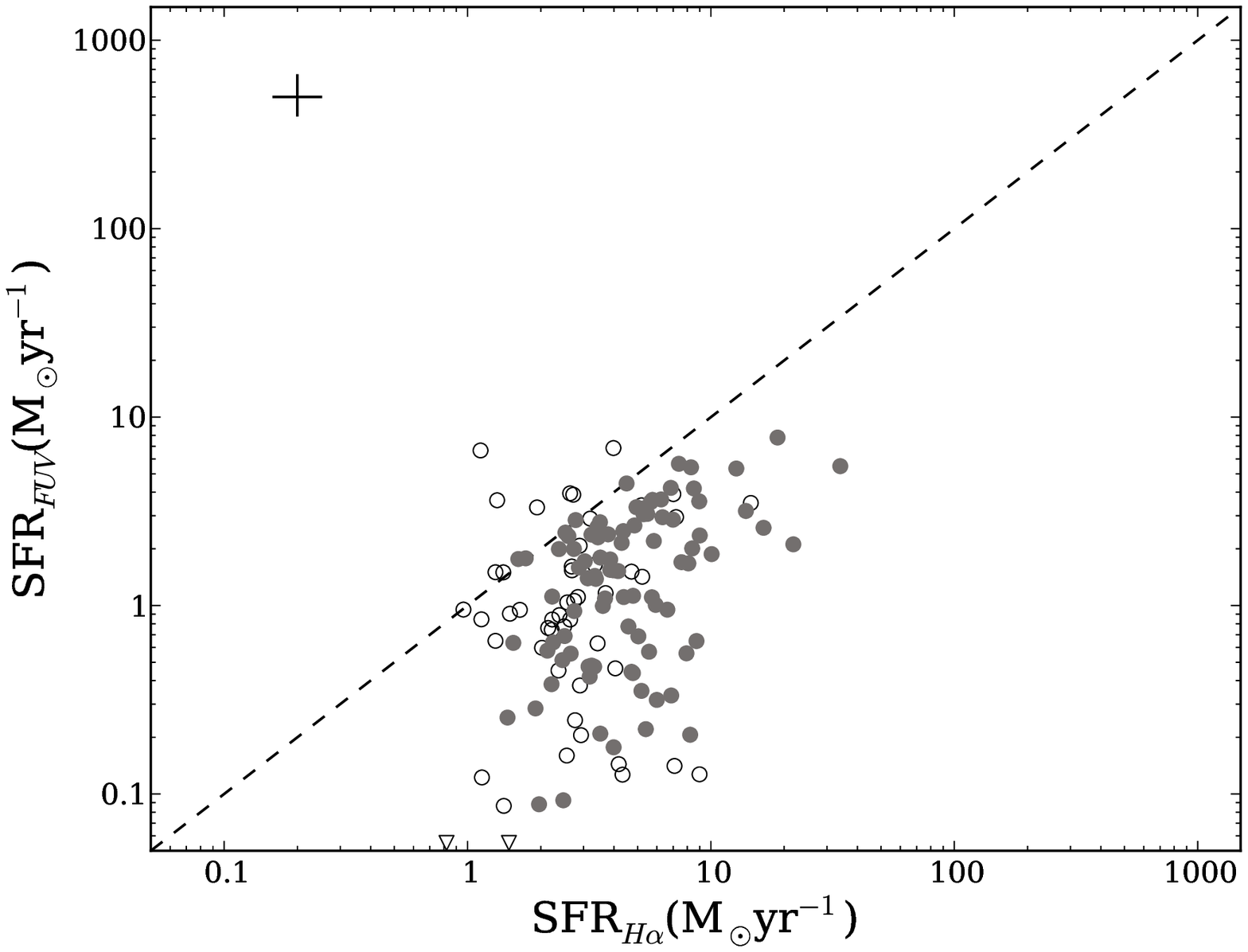}\\
\includegraphics[width=0.49\textwidth]{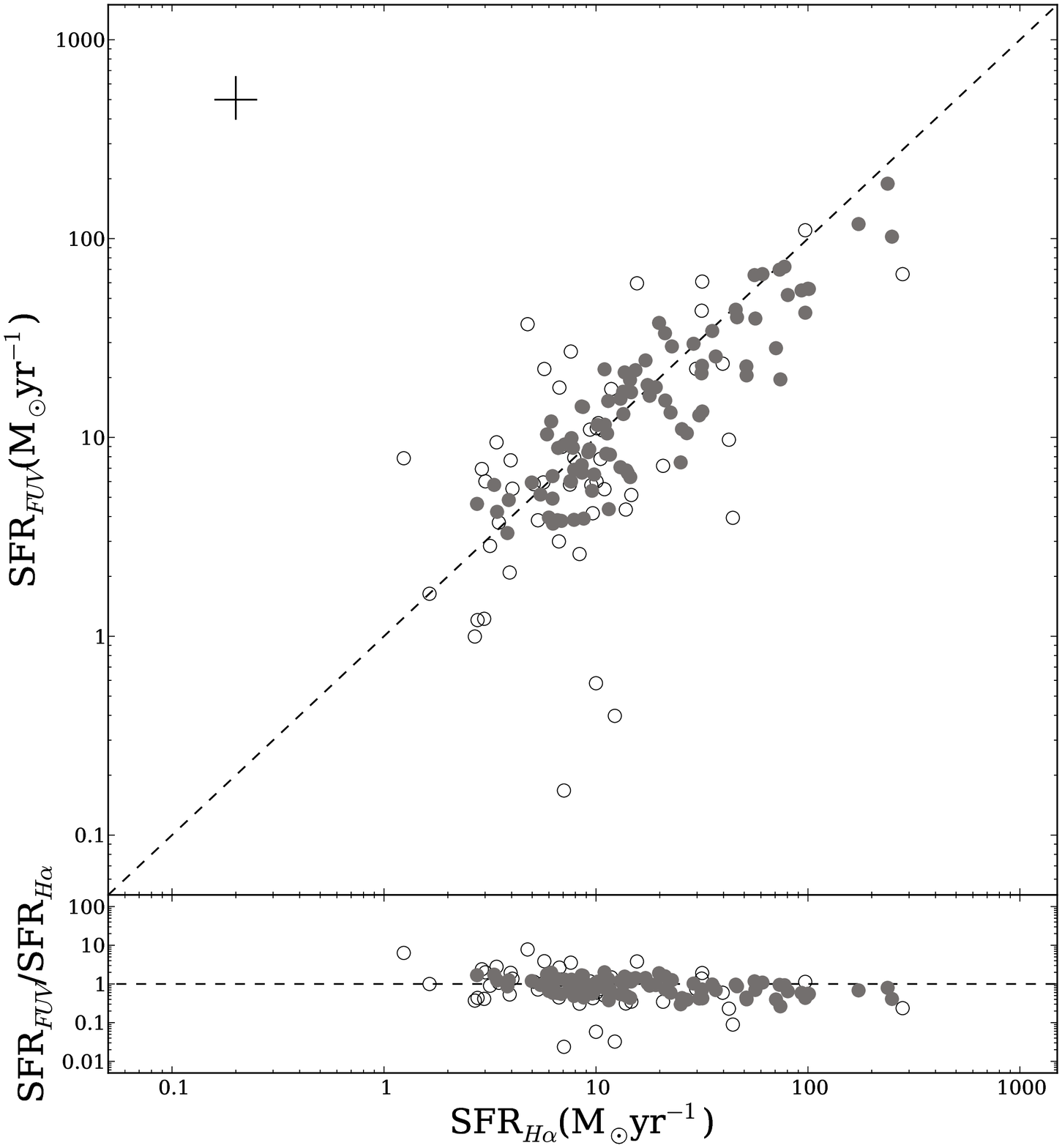}
\caption{\label{UV_Halpha} Comparison of SFRs inferred from H$\alpha$ luminosity and FUV luminosity. {\bf Top}: No extinction correction applied. {\bf Bottom}: Extinction correction applied to both tracers. Filled circles are objects confirmed by optical spectroscopy, whereas open circles are objects without spectroscopic confirmation. We show also the SFR$_{FUV}$/SFR$_{H\alpha}$ ratio versus SFR$_{H\alpha}$.}
\end{figure}

\begin{figure}[t]
\includegraphics[width=0.5\textwidth]{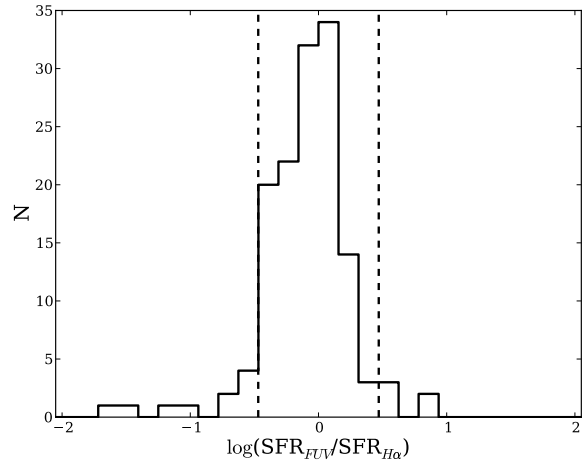}
\caption{\label{FUV_Halpha_ratio} Histogram of the ratio SFR$_{FUV}$/SFR$_{H\alpha}$ in logarithmic scale. Most of the objects concentrate around the unity ratio. There are also wings on both sides of the distribution, with some extreme cases where SFR is over(sub)-estimated up to 10-100. The tail of objects for which the UV sub-estimates the SFR is more extended than in the case of the H$\alpha$. The dashed lines encloses the objects whose SFRs agree within a factor of three.}
\end{figure}

In this section we compare the FUV derive SFR with that coming from the H$\alpha$ luminosity. The reader should note that in this section we use the FUV luminosity (1500\AA), which has not been used in the computation of $\gamma$ in the previous subsection, thus assuring the independence of the results.

The comparison between H$\alpha$ and FUV star formation rates is plotted in figure~\ref{UV_Halpha}. Top figure shows SFRs estimated without extinction corrections. Objects confirmed by optical spectroscopy are shown as filled circles whereas the objects lacking spectroscopic confirmation are shown as empty circles. The effect of the reddening is clearly visible as FUV SFRs are, in general, lower than those obtained through H$\alpha$. The median value and standard deviation for the FUV estimates are $\langle$SFR$_{FUV}\rangle$=1.5$^{+3.3}_{-0.9}$ M$_{\odot}$yr$^{-1}$, while for the H$\alpha$ line we find $\langle$SFR$_{H\alpha}\rangle$=3.5$^{+3.2}_{-1.7}$ M$_{\odot}$yr$^{-1}$. Objects not confirmed by spectroscopy show a higher dispersion, although compatible with that of the confirmed objects once a few outliers are removed.

In the bottom panel of figure~\ref{UV_Halpha} we show the effect of applying the extinction corrections. The SFR range spans considerably, going from 2-10 M$_{\odot}$yr$^{-1}$ when the effect of extinction is not corrected to 2-300 M$_{\odot}$yr$^{-1}$ when it is corrected. Estimations coming from both tracers now agree within a factor of 3. The statistical values are in this case: $\langle$SFR$_{FUV}\rangle$=10$^{+21}_{-7}$ M$_{\odot}$yr$^{ -1}$ and $\langle$SFR$_{H\alpha}\rangle$=11$^{+22}_{-7}$ M$_{\odot}$yr$^{ -1}$. The good agreement corroborates that our extinction corrections are working well and that these galaxies do not host star forming regions totally attenuated in the UV but visible in H$\alpha$, at least globally. There still can be regions totally obscured both in UV and in the optical, which will only arise in IR observations. We will explore this possibility in section \ref{ir}. 

To explore in more detail the differences between both tracers we study the SFR$_{FUV}$/SFR$_{H\alpha}$ ratio for each object. The median value is $\langle$SFR$_{FUV}$/SFR$_{H\alpha}\rangle$=0.89, which tells us that the H$\alpha$ line yields slightly higher values than the FUV for the star formation rate, although compatible considering the errors. As we use the FUV luminosity, which has not been used in the computation of $\gamma$, it is possible to have ratios below or above one. As the FUV is at shorter wavelength, the higher extinction could totally attenuate more regions than at higher wavelengths, thus underestimating the SFR. If we use SFRs obtained from 2800\AA\ instead, the ratio SFR$_{2800}$/SFR$_{H\alpha}$ becomes one, as this wavelength is in the regime used in the extinction law check. The number of objects is also different as objects used in the extinction section must have IR data. The distribution of ratios is shown in figure~\ref{FUV_Halpha_ratio}. Although the agreement is quite good, with 90\% of the objects within a factor of three, there are objects whose SFR is overestimated by H$\alpha$, with a few in the opposite case. If we consider only our spectroscopic confirmed sample we find almost the same results: $\langle$SFR$_{H\alpha}\rangle$=14$^{+23}_{-9}$ M$_{\odot}$yr$^{ -1}$ and $\langle$SFR$_{FUV}$/SFR$_{H\alpha}\rangle$=0.87.

The general agreement between both tracers is also found in the local Universe \citep{Salim07}. At z$\sim$2 \cite{Erb06} also compared these tracers for a sample of Lyman-break galaxies \citep{Steidel96,Steidel99} selected through the {\em U$_{n}$GR} criterion \citep{Adel04,Steidel04}. Their result shows good agreement between both tracers with a dispersion similar to that of our sample. At that same redshift, \cite{Yoshi10} find that both tracers are roughly consistent, although SFRs measured with H$\alpha$ are systematically larger by 0.3 dex. 

\subsubsection{Infrared}
\label{ir}

\begin{figure}[t]
\includegraphics[width=0.5\textwidth]{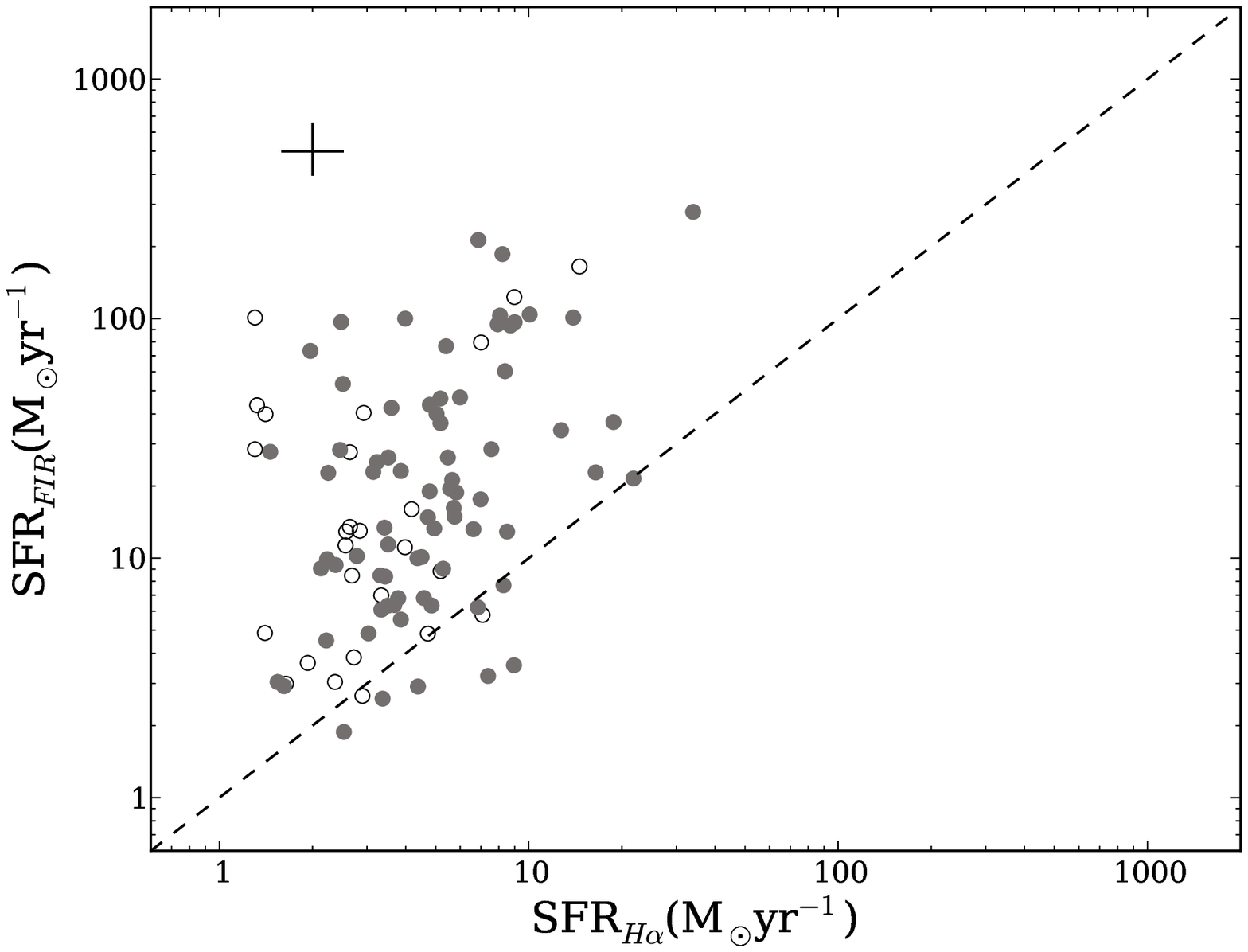}
\includegraphics[width=0.49\textwidth]{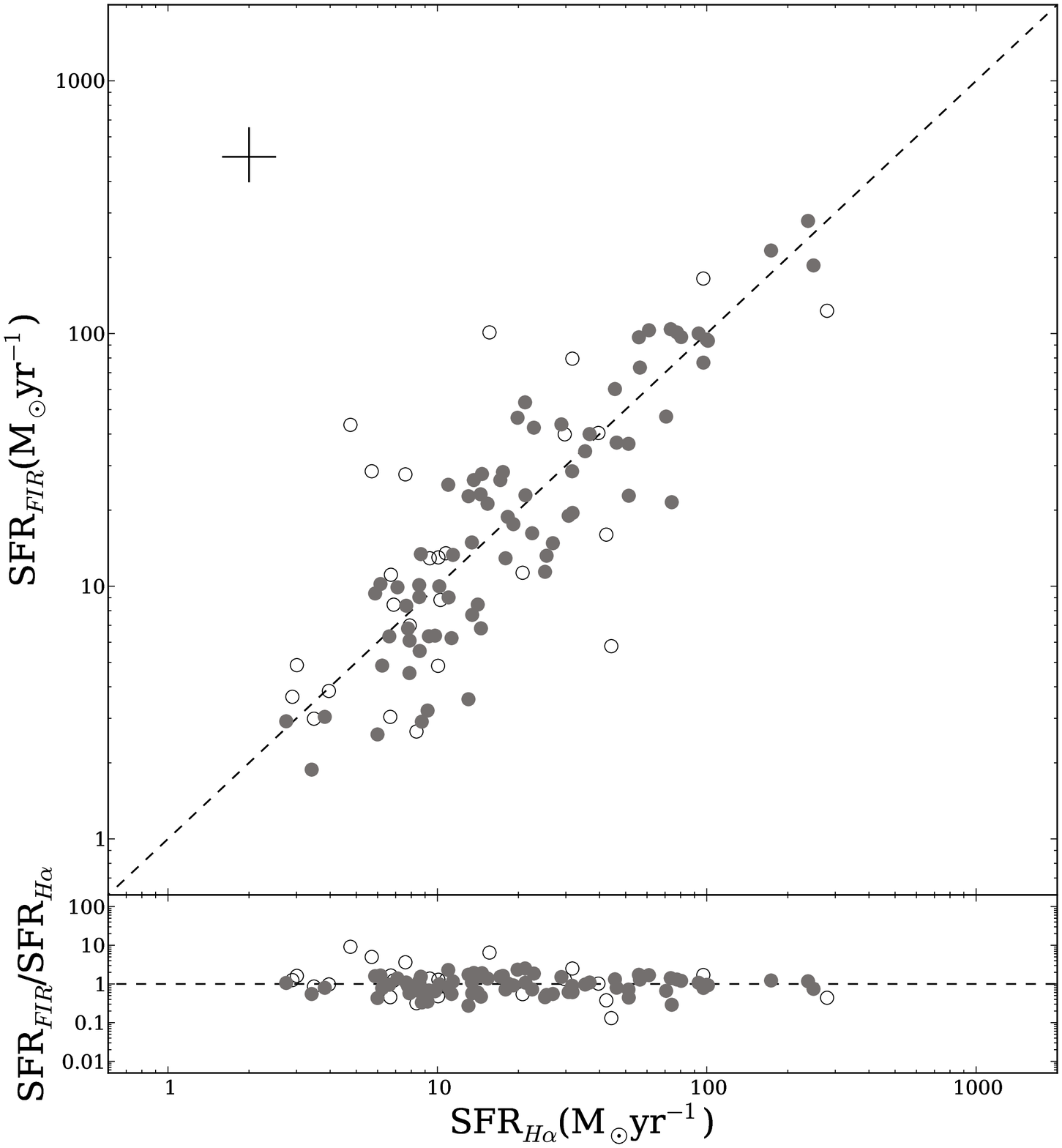}
\caption{\label{IR_Halpha} Comparison of SFRs inferred from H$\alpha$ luminosity and IR luminosity.  {\bf Top}: No extinction correction applied. {\bf Bottom}: Extinction correction applied to H$\alpha$. Filled circles are objects confirmed by optical spectroscopy, whereas open circles are objects without spectroscopic confirmation. We show also the SFR$_{IR}$/SFR$_{H\alpha}$ ratio versus SFR$_{H\alpha}$.}
\end{figure}

\begin{figure}[t]
\includegraphics[width=0.5\textwidth]{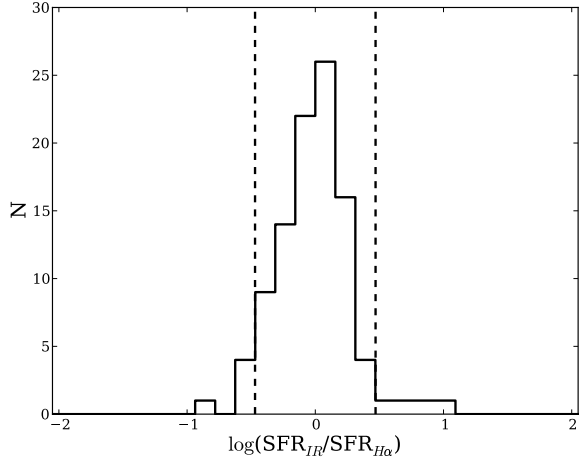}
\caption{\label{IR_Halpha_ratio} Histogram of the ratio SFR$_{IR}$/SFR$_{H\alpha}$ in logarithmic scale. Most of the objects concentrate around the unity ratio, though in general the IR estimates are higher than those obtained with H$\alpha$. The dashed lines encloses the objects whose SFRs agree within a factor of three.}
\end{figure}

Infrared SFRs are very interesting to check if  H$\alpha$ is losing substantial star formation due to dust attenuation. Top panel in figure~\ref{IR_Halpha} shows the comparison between both tracers before applying any extinction correction. Not surprisingly, H$\alpha$ sub-estimates systematically the SFR. There is a large scattering, which reflects the different attenuation that suffer each galaxy. It is worth noting that we only have infrared luminosities for the fraction of the sample detected with MIPS (100/140, 71\%). The IR limiting flux for a completeness of 80\% is 83$\mu$Jy in our surveyed fields, which translates into $\sim$10 M$_{\odot}$yr$^{ -1}$ for our redshift.

The median and standard deviation is $\langle$SFR$_{H\alpha}\rangle$=4.1$^{+3.7}_{-1.9}$ M$_{\odot}$yr$^{ -1}$ for H$\alpha$ while for the IR we obtain $\langle$SFR$_{IR}\rangle$=15$^{+34}_{-11}$ M$_{\odot}$yr$^{ -1}$. As in the case of UV, we do not observe systematic differences between spectroscopically confirmed and not confirmed objects.

Once the extinction corrections are applied (see bottom panel of figure~\ref{IR_Halpha}) we find $\langle$SFR$_{H\alpha}\rangle$=15$^{+29}_{-10}$ M$_{\odot}$yr$^{ -1}$, which agrees very well with the IR derive value. If we work with the ratios of SFRs we find that H$\alpha$ provides slightly higher estimates ($\langle$SFR$_{IR}$/SFR$_{H\alpha}\rangle$=0.95), but in agreement with IR estimates within uncertainties. In figure~\ref{IR_Halpha_ratio} we show the SFR$_{IR}$/SFR$_{H\alpha}$ distribution. In the figure it can be seen that H$\alpha$ estimates are systematically above the IR estimates. However, for 91\% of the objects, the star formation rates agree within a factor of 3. If we consider only our spectroscopic confirmed sample we find very similar results (see table~\ref{tab:comp}): $\langle$SFR$_{H\alpha}\rangle$=17$^{+30}_{-10}$ M$_{\odot}$yr$^{ -1}$ and $\langle$SFR$_{IR}$/SFR$_{H\alpha}\rangle$=0.96.

\subsection{Exploring the scatter}

It is interesting to explore the reasons that originate the discrepancies between different tracers. The H$\alpha$ line is only produced when the star forming region includes stars with masses above 10 M$_{\odot}$. Thus, only star forming regions aging less than 20 Myr are detectable through this line, since older regions would not have stars massive enough to photoionize the surrounding gas. There are other factors that could affect, such as metallicity, fraction of ionizing photons that escape, etc. This set of conditions do not hold for the other tracers, which have their own. Ultraviolet, for example, is more sensible to less massive stars, being more affected by the star formation history of the galaxy. Infrared is also affected by the star formation history as evolved stars can make a significant fraction of the infrared emission \citep{DaCunha08}. \cite{Ken09} find that 50\% of the infrared emission in the local galaxies within the SINGS sample come from evolved stars. At higher redshifts, \cite{Salim09} find for a sample with z$<$1.4 and with star formation (NUV-R$<$3.5), that the infrared flux fraction originated by intermediate and old stars can be as high as 60\%, with a typical value $\sim$40\%. 

\begin{figure}[t]
\includegraphics[width=0.5\textwidth]{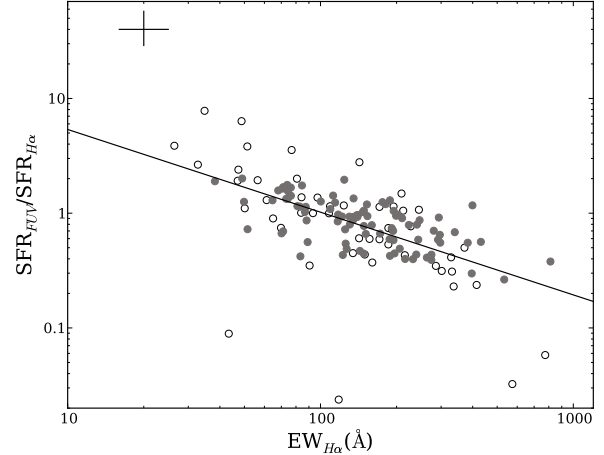}
\caption{\label{FUV_age} SFR$_{FUV}$/SFR$_{H\alpha}$ ratio versus EW(H$\alpha$). The line is the best linear fit to the data in logarithmic scale. The high EW(H$\alpha$) objects tend to have lower SFR$_{FUV}$/SFR$_{H\alpha}$ ratios whereas the lower EW(H$\alpha$) objects have tend to have higher ratios. }
\end{figure}

The calibration of the different tracers implies the assumption of a star formation history. This could lead to discrepancies when comparing different tracers. As a measure of star formation history we use the H$\alpha$ equivalent width. It is interesting to check if there is any systematic difference depending on age or star formation history. The equivalent width is defined as the quotient between the H$\alpha$ flux, which measures the relevance of the recently formed population, and the continuum flux under the line (by wavelength unit), which measures the contribution of the stars formed before. It is also related to the age of the star forming region \citep{PG03II}. In figure~\ref{FUV_age} we show the SFR$_{FUV}$/SFR$_{H\alpha}$ ratio versus the H$\alpha$ equivalent width. There exists an anti-correlation among both magnitudes, with the best fit given by:

\begin{equation}
\log (\frac{SFR_{FUV}}{SFR_{H\alpha}})= (1.45\pm0.64) - (0.72\pm0.29)  \times \log (EW_{H\alpha})
\end{equation} 

When the equivalent width is low, the weight of the young stars is less significant compared to that of the old population. In this case, the UV provides higher star formation rates than H$\alpha$, as it is more sensitive to older stars. As we move towards higher equivalent widths, the recently formed stars become more and more important over the old population. For the higher EW values the UV subestimates the SFRs compared to H$\alpha$. 

\begin{figure}[t]
\includegraphics[width=0.5\textwidth]{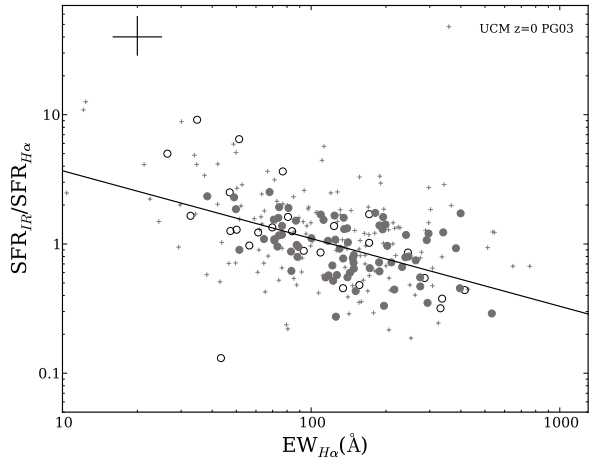}
\caption{\label{IR_age} SFR$_{IR}$/SFR$_{H\alpha}$ ratio versus EW(H$\alpha$). The line is the best linear fit to the data in logarithmic scale. The high EW(H$\alpha$) objects tend to have lower SFR$_{IR}$/SFR$_{H\alpha}$ ratios whereas the lower EW(H$\alpha$) objects have tend to have higher ratios. The crosses represent the UCM sample of local star forming galaxies, which show the same trend.}
\end{figure}

If we repeat this methodology regarding H$\alpha$ and the IR, we obtain similar results. An anti-correlation exists between the ratio SFR$_{IR}$/SFR$_{H\alpha}$ and EW(H$\alpha$) (see figure~\ref{IR_age}), which is described by the best linear fit as:

\begin{equation}
\log (\frac{SFR_{IR}}{SFR_{H\alpha}})= (1.09\pm0.76)-(0.52\pm0.35)\times \log (EW_{H\alpha})
\end{equation}

Again, as we move towards higher EW(H$\alpha$), i.e. to higher contributions of young stars, the SFR$_{IR}$/SFR$_{H\alpha}$ ratio decreases. We have also plotted the local UCM sample of star forming galaxies. These galaxies are located in the same region as the z$\sim$0.84 sample, so the effect of the star formation history on the SFR$_{IR}$/SFR$_{H\alpha}$ ratio is similar at both redshifts. In both cases, constraining to the spectroscopically confirmed sample, yields compatible results within errors. We note that the significance of both relations is not very high as errors are large, due to measurement errors, different star formation histories, etc.

\begin{figure}[t]
\includegraphics[width=0.5\textwidth]{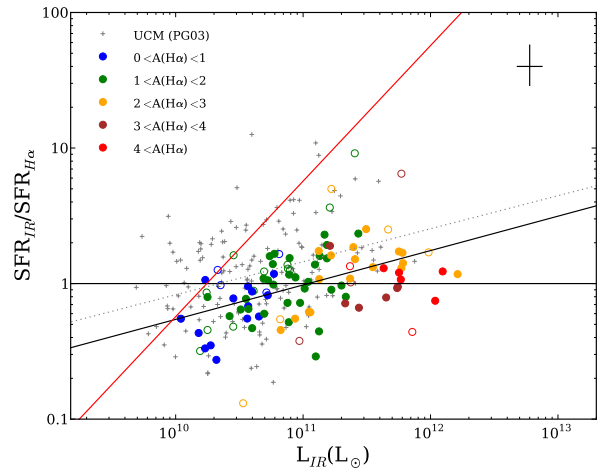}
\caption{\label{IR_RATIO} SFR$_{IR}$/SFR$_{H\alpha}$ ratio as a function of IR luminosity. The extinction is color coded as can be seen in the legend. The filled circles are confirmed by spectroscopy whereas the open circles are not. The thick line represents the best linear fit to the data (see text). The red line is our selection limit.  We could detect only objects below this line. The UCM local sample of star forming galaxies is represented by the crosses. The dotted gray line is the best linear fit to this sample. }
\end{figure}

Several authors \citep[][]{PG03T,Flores04,Hammer05} found a correlation between this ratio and the IR luminosity in the Local Universe. The authors argued that the more luminous is a galaxy in the infrared the more affected by extinction, to the point that the optical tracers could lose an important fraction of star formation, with some regions totally obscured by extinction. This would explain the sub-estimation of the SFR when measured by optical estimators, even when applying extinction corrections. At higher redshift \cite{Cardiel03} found a similar behavior for their sample at z=0.8. In figure~\ref{IR_RATIO} we represent the SFR$_{IR}$/SFR$_{H\alpha}$ ratio versus the infrared luminosity L$_{IR}$(8-1000$\mu$m). We find the same behavior reported by these authors: the H$\alpha$ estimator starts to sub-estimate the SFR (with respect to IR) when we move to higher L$_{IR}$(8-1000$\mu$m). The dependency it is not a selection effect as the limits of our sample would allow us to detect galaxies with higher and lower ratios (see figure). No dependency is found between the extinction of the objects (coded with different colors in the figure) and the degree of sub-estimation. One would expect some kind of dependency, as the obscured regions in the optical are visible in the IR, although it does not discard this scenario. 

The best linear fit, excluding the 5\% extreme values, is given by:

\begin{equation}
\log \left( \frac{SFR_{IR}}{SFR_{H\alpha}}\right) = - (2.61\pm2.17) + (0.23\pm0.19) \times {\log \left( \frac{L_{IR}}{L_{\odot}}\right) }  
\end{equation}

\noindent which can be seen in the figure as a thick line. 

This result agrees with previous results, with the IR providing higher SFRs as we move towards higher IR luminosities. We note that uncertainties are large and in this case, where we are representing x/y vs. x, a small correlation would arise from random scatter. However, the slope of such correlation is always below 0.1 within the range of our values, as we have checked generating random samples. We have plotted the UCM local sample as a reference, with its best linear fit as a dotted line in figure~\ref{IR_RATIO}. Both samples present very similar slopes when fitted, but there is an offset between them. For the UCM sample the IR starts to provide higher SFRs for lower IR luminosities than for the z$\sim$0.84 sample. If we consider that the change in the SFR$_{IR}$/SFR$_{H\alpha}$ ratio is due to the increment of H$\alpha$ luminosity totally obscured by dust, at higher IR luminosity more regions would be totally obscured by dust. Then, the difference between the local relation and that at z$\sim$0.84 presented in Figure~\ref{IR_RATIO} could be explained by a change in the number and size of star forming regions. These should be less numerous at z=0.84, although larger, given that SFRs are higher in the sample at z=0.84 than in the local sample. Thus, H$\alpha$ luminosity would be higher for each region and the dust would not totally attenuate that region. Only in very luminous galaxies in the IR with large amounts of dust it would be possible to totally attenuate some star forming regions in the visible. 

Thus, we have two possible effects that could explain the scatter when comparing SFRs coming from H$\alpha$ and IR: a) contribution of the evolved population and b) presence of star forming regions totally attenuated by dust.

However, figure~\ref{IR_RATIO} can be explained as well taking into account the contribution of the evolved population to the IR. We have shown that the SFR$_{IR}$/SFR$_{H\alpha}$ ratio increases with the IR luminosity and that there is an offset in that relation between the local Universe and z$\sim$0.84. But, the dependency of that ratio with the EW(H$\alpha$), i.e. with the weighted age, is independent of redshift. Then, we can consider that the same SFR$_{IR}$/SFR$_{H\alpha}$ ratio implies the same weighted age at z=0 and at z$\sim$0.84. On the other hand, galaxies with the same weighted age have higher infrared luminosity at z=0.84 than at z=0, as depicted in figure~\ref{IR_RATIO}. As their IR luminosity is higher, the star formation rate is higher, and, as the weighted age is the same, the underlying population has to be more luminous, i.e. more massive. This is in agreement with what can be expected from the {\em downsizing} \citep{Cowie96} scenario, where star formation moves from more massive galaxies at higher redshifts to less massive galaxies at lower redshifts (see section~\ref{sec:mass}).

Although the effect of age could explain the observed difference, probably both age and extreme attenuation of some regions are contributing. 

\section{Stellar masses}
\label{sec:mass}

The stellar mass is one of the most important properties of a galaxy, as it provides a robust measurement of the scale of the galaxy and it is also an indicator of the past star formation. The estimate of stellar mass is obtained from the best fitting template to the SED of each galaxy. The template provides mass-to-light ratios for each observed band and a stellar mass is computed for each one. The final value is the average of the values obtained for each observed band, being the associated error the standard deviation of the distribution of stellar masses. The results are more reliable than those obtained through a single mass-to-light ratio, as it is less sensitive to the star formation history or errors in photometry or templates. For more details on the procedure see \cite{PG08} and  \cite{Barro11b}.

Stellar masses were obtained with the PEGASE 2.0 \citep{Fioc97} stellar population synthesis models, a \cite{Salpeter55} initial mass function (IMF) and the \cite{Cal00} extinction law. Different stellar population models \citep[][Charlot \& Bruzual 2009]{Bruzual03,Maraston05}, IMFs \citep{Kroupa01,Chabrier03} or extinction laws will provide different estimates. The models used here predict the largest stellar masses, although all models are roughly consistent within a factor of 2. For a detailed comparison between the different models we refer the reader to  \cite{Barro11b}.

\begin{figure}[t]
\includegraphics[width=0.5\textwidth]{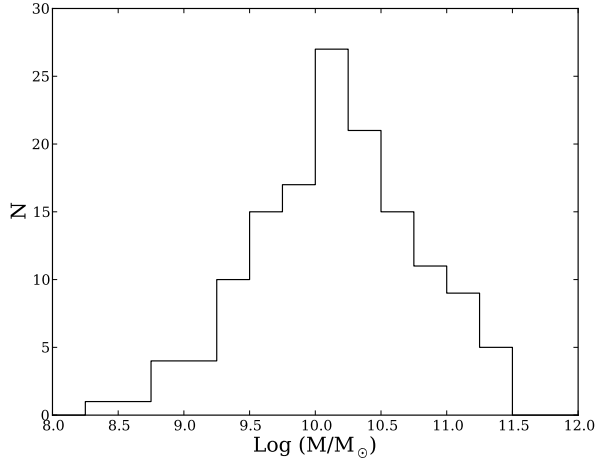}
\caption{\label{MASS_HIST} Histogram of stellar masses for our sample. The completeness falls below 10$^{10}$M$_{\odot}$. }
\end{figure}

In figure~\ref{MASS_HIST} we present the histogram of masses for our sample. The median and standard deviation for the distribution are M$_{\star}$=1.4$^{+4.6}_{-1.1} \times 10^{10}$M$_{\odot}$. At the same redshift, the typical mass found by \cite{PG08} for an IRAC selected sample is M$^{*}_{\star}$=1.6 $\times10^{11} $M$_{\odot}$. The typical mass of an star-forming galaxy (to the limit of our sample) is ten times lower than the typical mass of the global population of galaxies. \cite{Sobral10} (hereafter S10) find a typical mass M$_{\star}$=2.25$\times 10^{10} $M$_{\odot}$ (after scaling from a Chabrier to a Salpeter IMF) for their HiZELS sample at z=0.84, in very good agreement with ours, given that both limiting fluxes are very similar.

The loss of the low mass population of star forming galaxies is clear in the histogram, with the number of galaxies starting to decrease below $\sim$ 10$^{10}$ M$_{\odot}$. This effect is produced by the limiting line flux reached in our selection process, given the correlation between stellar mass and star formation found in the Local Universe \citep[][hereafter B04]{Brinchmann04} and at higher redshifts up to z$\sim$6 \citep{Noeske07a,Elbaz07,Daddi07,Stark09}. 

\begin{figure*}[t]
\includegraphics[width=1.0\textwidth]{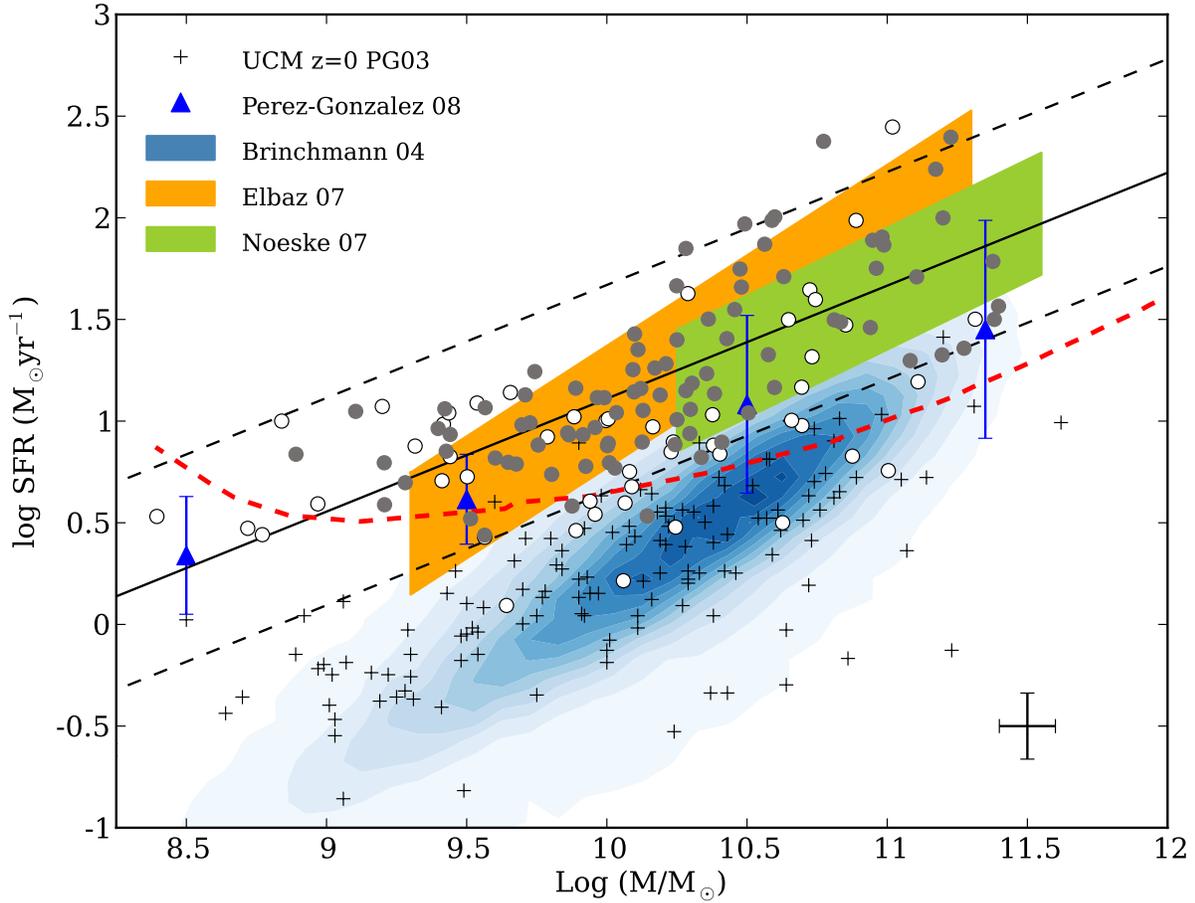}
\caption{\label{MASS_SFR} H$\alpha$ star formation rate versus mass. The circles represent our sample: filled for the spectroscopic confirmations and open for the remainder. The red dashed line represents the 50\% completeness level. The best fit within mass completeness limits is shown as a solid line, with the dashed lines enclosing 90\% of the data. The filled region are the best fits obtained by \cite{Elbaz07} and \cite{Noeske07a} \citep[compiled in][]{Dutton10}. The mass selected sample values at z=0.8-1.0 obtained by \cite[][PG08]{PG08} are represented by the blue crosses. The UCM (crosses) and SDSS (blue colormap) local samples are also shown.}
\end{figure*}

The SFR-M$_{\star}$ correlation for our sample is shown in figure~\ref{MASS_SFR}, where the H$\alpha$ SFR versus the stellar mass is represented. The completeness level (red dashed line) is estimated performing simulations of the whole selection process and taking into account the extinction. This process involves several steps which are explained in detail in appendix~\ref{App:completeness}. The SFR-M$_{\star}$ relations obtained from \cite{Dutton10}, from the samples presented in \cite{Noeske07a} at z$\sim$0.8 and in \cite{Elbaz07} at z$\sim$1.0, in good agreement with our sample, are overplotted. In addition, the UCM local sample of star forming galaxies \citep{PG03II} and the SDSS DR4 galaxies classified as star forming in B04 are also shown. The slope for these samples are similar to ours, although the SDSS is steeper. There is a shift in SFR between the local ones and that at z$\sim$0.84. For a given mass the sample at z$\sim$0.84 presents higher ($\sim\times$5.5) star formation rates than the local sample. Another way to see it is that for a given star formation rate, the galaxies in the past were less massive than local galaxies. This difference between the local Universe and z$\sim$0.84 clearly shows that star formation changes as the Universe evolves. Contrary to our result, S10 do not find any relation between SFR and stellar mass. This is intriguing as the selection technique and line flux reached are very similar in both surveys. 

In this type of comparisons between our sample and the local samples one could think that we are just watching the tail of the distribution for the z$\sim$0.84 sample, which lead us to the wrong conclusion that an evolutionary effect is present. In \citetalias{Villar08} we computed the H$\alpha$ luminosity function and we also estimated the completeness limit. The results showed that we start to lose a substantial fraction of objects one order of magnitude below L$^{*}$($H\alpha$) (obtained from the LF fit to a Schechter function). We were 50\% complete for log L(H$\alpha$) $>$ 42.0, which in terms of L$^{*}$(H$\alpha$) is log(L($H\alpha$)/L$^{*}$($H\alpha$)) $\sim$ -1.0. Therefore, we conclude that we are not observing the rare galaxies in the tail of the distribution. On the other hand, the UCM sample is considered complete down to log L(H$\alpha$) $\sim$ 40.7, which in terms of L$^{*}$(H$\alpha$) (for this redshift) is log (L($H\alpha$)/L$^{*}$($H\alpha$)) $\sim$ -1.2. Thus we are reaching very similar H$\alpha$ luminosities for both samples in terms of L$^{*}$(H$\alpha$). Another argument supporting this conclusion is that we detect essentially the same population as samples selected in UV or IR at the same redshift \citepalias[see][]{Villar08}

It is also interesting to compare with a sample purely selected in stellar mass, as that presented in \cite{PG08}. The median values of this sample at z=0.8-1.0 fall very close to our best fit, being compatible within errors. However, the difference between our best fit and the median values of that sample increases as we move to higher masses, presenting the mass selected sample lower SFRs. Although this trend is very weak and, given the errors, could even not be present, is consistent with the {\em downsizing} scenario \citep{Cowie96}: the fraction of galaxies with very low star formation seem to increase as we move to higher masses, decreasing the median SFR in the mass selected sample. At low masses the effect is less pronounced as galaxies are undergoing strong star formation episodes and are still very active in general.

The SFR-M$_{\star}$ correlation allows us to estimate the mass range in which we can consider our sample unbiased. The cut between the SFR 50\% completeness level and the lower envelope in the SFR-M$_{\star}$ distribution gives us the stellar mass range within we can consider our sample free of biases. The lower envelope is the best linear fit shifted downwards to enclose 95\% of the data confirmed by spectroscopy (90\% within lower and upper envelope). In practice it is an iterative process: first we fit the data above and below initial mass limits, then we find the new mass limits in the intersection between the lower envelope and the completeness curve, and the procedure is repeated until the mass limits used in the fit and the ones obtained converge. This gives us a lower limit of $\sim$10$^{10}$M$_{\odot}$ and no upper limit, indicating that we are limited by explored volume on the upper side, as we do not detect any galaxy above $\sim$3$\times$10$^{11}$M$_{\odot}$. As mentioned before, the derived correlation is in good agreement with \cite{Noeske07a} and \cite{Elbaz07}, both at a similar redshift. Moreover, we find that the scatter is $\sim$0.3 dex, which is the typical value found in other studies and seems to be almost constant with redshift \citep{Dutton10}. However, the lower envelope and the 50\% completeness limit are close, and thus it is still possible that the observed correlation is produced by the selection effect. To rule out this possibility we simulate fake samples of galaxies following different SFR-M$^{*}$ relations. First we simulate a population of galaxies following the linear relation derived from our sample. We assign to each galaxy a random stellar mass. Given this stellar mass we compute the SFR with the linear relation derived from the real sample, adding gaussian noise to simulate the scatter ($\sigma$=0.3). Once we have this fake population we check if the galaxies would be detected considering the completeness curves (measured for different completeness levels) and the volume sampled. After repeating the fitting process considering the completeness curve as well as the lower envelope, we obtain that the results are in good agreement with the input, with relative errors within 10\% for the slope and 20\% for the constant term. The same process is repeated 20 times to avoid biases due to rare distributions.  When a flatter relation (20\% less steep) is used we find similar results, although in some cases the lower envelope is so low that it does not cut the completeness curve and no measure can be obtained. Thus, given the SFR-M$_{\star}$ relation and the limits of our sample we can be confident that we will not introduce any substantial bias when inferring mass related properties using the sample within those mass limits.

\begin{figure}[t]
\includegraphics[width=0.5\textwidth]{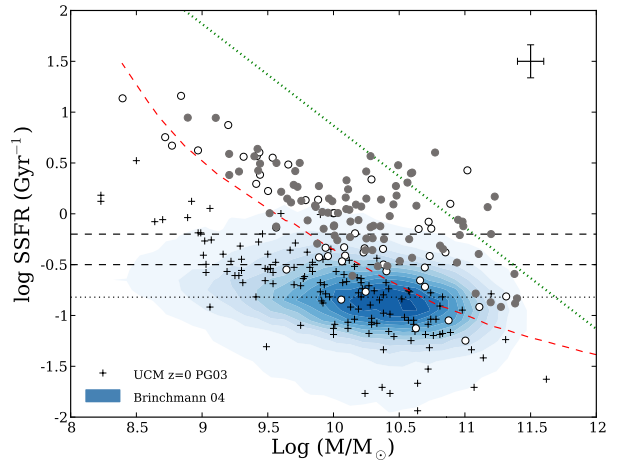}
\caption{\label{MASS_SSFR} Specific H$\alpha$ star formation rate versus mass. The circles represent our sample: filled for the spectroscopic confirmations and open for the remainder. Our 50\% completeness level is represented by the red dashed line. The diagonal dotted line is the place that would occupy the objects with L(H$\alpha$)=L$^{*}$(H$\alpha$). The crosses represent the UCM sample of local star forming galaxies. The color map shows the values for the SDSS sample. The horizontal dashed lines indicate the values of sSFR at which {\em b}=1 and {\em b}=2 at z=0.84. The dotted lines are the same but for the local Universe and {\em b}=1. The line for {\em b}=2 coincide with the {\em b}=1 line for z=0.84.}
\end{figure}

Figure~\ref{MASS_SFR} could be interpreted as going against {\em downsizing}, as galaxies at z$\sim$0.84 have higher SFRs than local ones, independently of their stellar mass. The key concept here is not the absolute SFR, but the sSFR, which is the SFR per unit of stellar mass, and thus is a good indicator of the impact that the star formation has in the galaxy. In figure~\ref{MASS_SSFR} we represent the sSFR for our sample and for the SDSS and UCM local samples. Now it is clear the change in star formation as we move towards the local Universe, as well as the shift in star formation from more massive to less massive galaxies, considering the evolutionary impact of the SFR processes.

There exists an anti-correlation between the sSFR and the stellar mass, evidencing that star formation processes have a higher impact on less massive galaxies. This trend is also present in the star forming galaxies of the local Universe, with a similar slope in the case of the UCM sample, although shifted in the sSFR axis to lower values, indicating that the star formation is less important. Most of our sample fall below the line (green dotted line) where the L$^{*}$(H$\alpha$) galaxies would lie at that redshift, which shows that we are not missing the general population. Previous determinations \citep[e.g][S10]{PG05,Noeske07b,Rodighiero10} already found this relation but discrepancies arise regarding the slope of the correlation. \cite{Rodighiero10}, through IR SFR estimations, found a flatter relation compared to \cite{Noeske07b}, who used UV-optical SFR estimators. On the contrary, S10 found a much steeper relationship, with a slope $\sim\ $-1, given that these authors do not find any correlation between SFR and mass. 
 
The slope for our sample (computed only for objects within our mass limits) is $\beta$=-0.4$\pm$0.1, which is in good agreement with that of \cite{Noeske07a}, but steeper than the value obtained by \cite{Rodighiero10} ($\beta$=-0.28) in the redshift range 0.5 $<$ z $<$ 1.0. This discrepancy could come from the different selection criteria. While in this paper galaxies are selected purely by star formation, \citeauthor{Rodighiero10} analyzed a sample selected in mass through the IRAC 4.5$\mu$m band, with a color cut to avoid the inclusion of red sequence galaxies, although they try to recover dusty starbursts that might fall in the red sequence. In this work we have not excluded any galaxy by its color since, as we have shown, most of the red-sequence galaxies were in fact dusty star forming galaxies. In line with this hypothesis, \cite{Karim10} found a similar slope to ours ($\beta$=-0.38), using a sample of star forming galaxies selected from a mass selected sample in the IRAC 3.6$\mu$m band. However, in this case the authors classify the objects as star-forming if they belong to the blue cloud, once the attenuation has been removed. \cite{Gilbank11} have found very recently a very similar value for the slope ($\beta$=-0.42) using a spectroscopic sample with [OII] SFRs around z$\sim$1.

The birth rate parameter {\em b} is linked to the sSFR:

\begin{equation}
b=\frac{SFR}{\langle SFR \rangle} = SFR \frac{t_{f}}{2 \times M_{*}} 
\end{equation}

\noindent where ${\langle SFR \rangle}$ is the average star formation in the whole history of the galaxy, or, in other words, the average SFR that would have produced the current stellar mass. Thus, we can obtain this average star formation dividing the stellar mass $M_{*}$ by the elapsed time since the galaxy formed {\em t$_{f}$}. The factor two takes into account the stellar mass returned to the interstellar medium. The parameter depends on the choice of the beginning of star formation, but it is still very interesting when comparing populations at different redshifts. In this work we set this initial time at the beginning of the Universe, t$_{f}$=6.5 Gyr for z=0.84 and t$_{f}$=13.4 Gyr for the local Universe. A galaxy with a value of {\em b} higher than one tell us that the current star formation is more intense than the average star formation in the past. A value of two indicates a specially intense star formation episode.

Most of our galaxy sample presents values of {\em b} higher than one, with 85 \% (119/140) with b$>$1 and 66 \% (92/140) with b$>$2. In the local Universe the SDSS and UCM samples show a very different scenario. Half the SDSS sample (48\%) have {\em b} values over one and only 16\% above two. The UCM sample presents similar results, with 58\% of the sample with {\em b}$>$1 and 25\% with {\em b}$>$2. If we confine the analysis to the most massive galaxies (with stellar masses above 10$^{11}$M$_{\odot}$) within our sample, 36\% (5/14) of these have {\em b} values above one, and 29\% (4/14) have values over two. Regarding the local samples, the proportions are very different, with 20\% in the case of the SDSS and 14\% in the case of the UCM  with {\em b} over one, and only 3\% with {\em b} over 2 within the SDSS sample and none in the UCM sample. This is a direct evidence of {\em downsizing}, as the fraction of most massive galaxies with intense ({\em b}$>$1) or very intense ({\em b}$>$2) star formation was higher in the past, and it has reduced dramatically from that epoch to the present.

\subsection{Quenching Mass}
 
The star formation mass relation holds up until a certain mass, above which it no longer holds and the star formation drops sharply (see for example B04). Galaxies above this mass are considered quiescent, as star formation processes are no longer the main drivers of its evolution and they move to the red sequence. Therefore, this mass defines an upper limit to the stellar mass of the galaxies actively forming stars. \cite{Bundy06} (hereafter B06) found that this {\em quenching} mass evolves with redshift, increasing as we move to higher redshift, as expected in the {\em downsizing} scenario. In this work, we use our H$\alpha$ selected samples to study this {\em quenching} mass evolution, which our star forming samples also reflect. The decrease of sSFR with mass implies that galaxies will eventually reach a mass over which the star formation processes will be very low and we can consider them quiescent. The observed shift between both trends leads to a different {\em quenching} masses, being lower that at the local Universe.   

With our data it is possible to estimate an upper limit for the {\em quenching} mass. For the sake of clarity we are going to use the {\em doubling} time $t_{d}$, which is analogous to the sSFR and is defined as:

\begin{equation}
t_{d}=\frac{M_{*}}{SFR\ (1-R)}=\frac{1}{SSFR\ (1-R)} 
\end{equation}

\noindent where {\em R} is the fraction of mass returned to the interstellar medium which is generally assumed to be $\sim$0.5 \citep{Bell03}.  

\begin{figure}[t]
\includegraphics[width=0.5\textwidth]{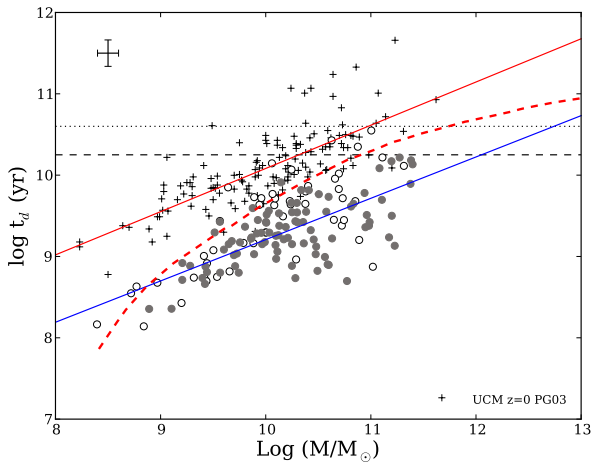}
\caption{\label{MASS_TD} Doubling time {\em t$_{d}$} versus stellar mass. The circles represent our sample: filled for the spectroscopic confirmations and open for the remainder. The 50\% completeness level is represented by the red dashed line. The crosses represent the UCM local sample of star forming galaxies. The dashed horizontal line indicates the doubling time at which we consider a galaxy at z=0.84 as quiescent. The dotted horizontal line represents the same but for the local Universe. Best linear fits computed for both samples are also shown.}
\end{figure}

The {\em doubling} time tells us how long will take for that galaxy to duplicate its stellar mass if its current star formation stays constant. Galaxies with a large {\em doubling} time will evolve slowly whereas galaxies with a small one will evolve quickly. {\em Doubling} times versus mass for our sample and the local UCM sample are shown in figure~\ref{MASS_TD}. In order to estimate the {\em quenching} mass we define a galaxy as quiescent if its {\em doubling} time is higher than what we define as {\em quenching} time: $t_{Q}=3 \times t_{H}$, where $t_{H}$ is the Hubble time. 
To obtain the typical mass which corresponds to the {\em quenching} time we performed several steps. First, we simulated 1000 realizations of our sample, varying randomly the values of SFR and mass within twice the errors, i.e., each object will have values randomly distributed in the intervals [M-2$\Delta$M, M+2$\Delta$M] and [SFR$_{H\alpha}$-2$\Delta$SFR$_{H\alpha}$, SFR$_{H\alpha}$+2$\Delta$SFR$_{H\alpha}$]. Second, we do a linear fit of $t_{d}$ versus mass only with the objects whose simulated mass fall above our mass limit. For each of these fits we compute the {\em quenching} mass as the mass at which the {\em doubling} time $t_{d}$ is equal to the {\em quenching} time t$_{Q}$. The final {\em quenching} mass M$_{Q}$ is the median of the whole distribution of {\em quenching masses}, with the error determined by the standard deviation of the distribution. The same process has been followed for the UCM sample. 

We obtain that M$_{Q}$=1.0$^{+0.6}_{-0.4}\times$10$^{12}\, M_{\odot}$ (log~(M$_{Q}$/M$_{\odot}$)=12.0$\pm$0.2) for the z$\sim$0.84 sample and M$_{Q}$=7.9$^{+1.9}_{-1.5}\times$10$^{10}\, M_{\odot}$ (log (M$_{Q}$/M$_{\odot}$)=10.9$\pm$0.1) for the local sample. If we consider only the spectroscopically confirmed sample we obtain log~(M$_{Q}$/M$_{\odot}$)=12.2$\pm$0.2, slightly higher although compatible within errors. In the case of z$\sim$0.84 the {\em quenching} mass is outside the range of masses detected, given the limit on the detection of massive galaxies imposed by sampled volume and the equivalent width limit of the survey, which prevents us from selecting objects with lower sSFRs. These masses are upper limits, given that at high stellar masses the correlation between {\em doubling} time and mass will break as a consequence of {\em quenching}. Galaxies with higher t$_{d}$ than predicted by the correlation will appear as the {\em quenching} takes over, possibly lowering the average {\em quenching} mass, specially in the case of z$\sim$0.84, where no galaxies around M$_{Q}$ have been detected. In order to detect these galaxies it would be necessary to survey larger volumes. In addition, the simulations (see section~\ref{sec:mass}) show that we may overestimate the quenching mass at z$\sim$0.84 by $\sim$0.1 dex, due to the completeness limits. However, these simulations also shows that we would be able to detect quenching masses $\sim$0.5 dex lower (SFR-M$_{*}$ slope 20\% lower), with a similar dispersion.

\begin{figure}[t]
\includegraphics[width=0.5\textwidth]{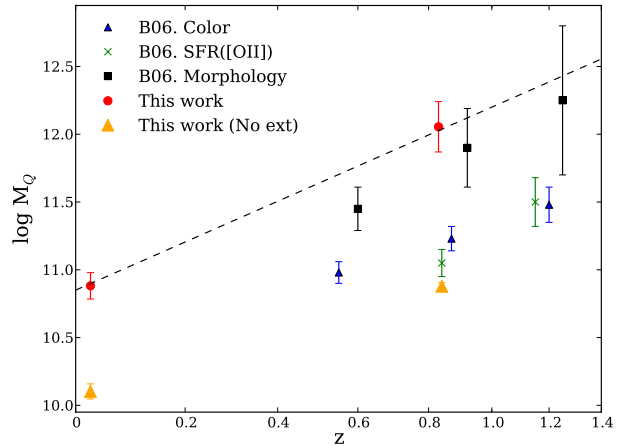}
\caption{\label{MQ_EVO} Evolution of the quenching mass limit M$_{Q}$. Red circles represent the results obtained in this work for the H$\alpha$ selected samples at z$\sim$0.84 and the local Universe. The orange circles are the estimated quenching masses when no extinction correction is considered. The rest of the points correspond to the B06 work according to the different criteria employed: black squares for morphology, green crosses for [OII]$\lambda$3727\AA\ SFRs and blue triangles for the (U-B) color. }
\end{figure}

Our {\em quenching} mass estimation for the local Universe is in very good agreement with the stellar mass ($\sim$7$\times$10$^{10}$ M$_{\odot}$; scaled from a Kroupa to a Salpeter IMF) above which \cite{Kauff03b} found a rapid increase in the fraction of galaxies with old population in the SDSS local sample. This change, detected by a transition from lower values of D$_{n}$(4000) to higher values, is also seen as a change in the slope of the $\mu_{*}$-M$_{*}$ (surface stellar mass density versus stellar mass)  correlation. Our result at z$\sim$0.84 is higher than those estimated by B06 at a similar redshift. In their work, they used three different approaches: morphology, U-B color and SFRs derived from [OII] equivalent width. Through the morphology criterion they obtained M$_{Q}\sim$8$\times$10$^{11}$M$_{\odot}$ whereas both color and SFR criteria provide lower masses $\sim$10$^{11}$ M$_{\odot}$ (see figure~\ref{MQ_EVO}). We have scaled B06 masses from the Chabrier IMF to the Salpeter IMF used in this work adding 0.25 dex. Our result is in good agreement with their morphology based estimation but it is higher than those based in color or SFR. B06 attributed this difference to a longer timescale in the processes that transform late types into early types. However our value is solely based in star formation and no morphology considerations have been done. One of the caveats of their SFR and color measurements is that extinction was not corrected. Therefore, dusty starbursts would appear redder and with lower SFRs, as they would be classified as red or non-star forming galaxies, which translates in lower quenching masses. If we estimate again M$_{Q}$ for our H$\alpha$ selected samples, but this time without applying the extinction correction, we obtain lower values:  M$_{Q}$=7.6$^{+1.7}_{-1.4}\times$10$^{10}\, M_{\odot}$ log~M$_{Q}$/M$_{\odot}$=10.88$\pm$0.09) for the z$\sim$0.84 sample and M$_{Q}$=1.3$^{+0.2}_{-0.2}\times$10$^{10}\, M_{\odot}$ (log~M$_{Q}$/M$_{\odot}$=10.13$\pm$0.05) for the local sample. The effect of the extinction is very high, and is enough to account for the difference between B06 morphology and color/SFR estimations. Our result is also consistent with the work by \cite{PG08}, where they found that at z$\sim$0.8 all the stellar mass has been already assembled for objects more massive than M$_{\star}$=10$^{12}$M$_{\odot}$ and almost fully assembled for objects with stellar mass in the range 10$^{11.7}$-10$^{12}$M$_{\odot}$.

The {\em quenching} masses estimated through this method rely on the definition of t$_{Q}$, however, independently of this parameter, we find a strong evolution between the local Universe and z$\sim$0.84. In the local Universe, galaxies with mass higher than 10$^{11}$M$_{\odot}$ are quiescent and their evolution is limited to interactions with other galaxies via dry mergers (or other processes not involving massive star formation), whereas at z$\sim$0.84, galaxies with mass in the range $\sim$10$^{11}-\sim$10$^{12}$M$_{\odot}$ are still under strong star formation processes. This is, again, in good agreement with the {\em downsizing} scenario. Despite we do not have enough data to constrain the M$_{Q}$ evolution with redshift we find that our results are compatible with the parametrization given by B06, i.e M$_{Q}\propto$(1+z)$^{4.5}$. The added value is that we have extended the redshift baseline to the local Universe, using samples selected uniformly.  

\begin{deluxetable}{ccc}
\tablecaption{Comparison between total and confirmed samples.}
\tablehead{
\colhead{Property} & \colhead{Total Sample} & \colhead{Confirmed Sample} \\
\colhead{(1)} & \colhead{(2)} & \colhead{(3)}
}
\startdata
$\gamma$ Calzetti et al.(2000) & 0.55 & 0.56\\
$\langle$SFR$_{H\alpha}\rangle$ & 11$^{+22}_{-7}$ M$_{\odot}$yr$^{ -1}$ & 14$^{+23}_{-9}$ M$_{\odot}$yr$^{ -1}$\\
$\langle$SFR$_{FUV}$/SFR$_{H\alpha}\rangle$ & 0.89 & 0.87\\
$\langle$SFR$_{IR}$/SFR$_{H\alpha}\rangle$ & 0.95 & 0.96\\
log~(M$_{Q}$/M$_{\odot}$) & 12.0$\pm$0.2 & 12.2$\pm$0.2\\
\enddata
\tablecomments{(1) Measured property (2) Value obtained using the whole sample (3) Value obtained using the sample confirmed with optical spectroscopy}
\label{tab:comp}
\end{deluxetable}

\section{Summary and Conclusions}
\label{sec:conclusions}

In this work we have analyzed the properties of an H$\alpha$ selected sample of star-forming galaxies at z$\sim$0.84, focusing on the star formation and stellar mass.  

We have discarded the AGN contaminants through two criteria: X-ray luminosities and IRAC colors. We find seven counterparts in X-ray, though three of them present very low fluxes, compatible with being originated by star formation processes. Thus, we only discard the four objects with fluxes high enough to have an AGN origin. Using IRAC colors we find another ten objects (one of them already detected in X-rays) that fulfill the criterion to be considered AGN. A total of thirteen objects are finally discarded.

The objects of our sample present a median $M_{B}$=-20.5$\pm$0.9$^{m}$, brighter by more than one magnitude than the UCM local sample of star forming galaxies. Most of the galaxies belong to the blue sequence with a small fraction of objects in the green valley and the red sequence. Once the extinction corrections are applied all these red objects except two move to the blue sequence, unveiling their dusty nature.

A check on the extinction law reveals that the \cite{Cal00} extinction law is appropriate for our objects, but with E(B-V)$_{stars}$=0.55 E(B-V)$_{gas}$.

The H$\alpha$ SFR, without applying the extinction correction, presents values in the range 2--10 M$_{\odot}$. We have also estimated SFRs with FUV and IR. In the first case, the non-extinction corrected FUV underestimates the SFR with respect to H$\alpha$. The opposite case is given for the IR, which overestimates the SFR with respect to H$\alpha$. These discrepancies are mainly driven by the extinction. Once we apply the extinction correction to both FUV and H$\alpha$ estimations, all SFR tracers agree within a factor of three and the highest SFRs reach several hundreds solar masses per year.

The scattering between the different tracers are correlated with the H$\alpha$ equivalent width. This  can be explained through the different weighted age of the objects (which is related to the EW) and the fact that FUV an IR SFRs are sensitive to a longer time range than H$\alpha$, being more affected by older populations. 

We have estimated stellar masses for our sample, finding that the median value is M$_{\star}$=1.4$^{+4.6}_{-1.1} \times 10^{10} $M$_{\odot}$, in good agreement with the result obtained by S10 for a sample selected with similar criteria as ours. The typical mass found by \cite{PG08} for an IRAC selected sample at this redshift is ten times higher than this value.

Our sample shows a trend between SFR and stellar mass. The slope of this trend is in good agreement with the value obtained by \cite{Noeske07b} and is flatter than the \cite{Elbaz07} result. The trend is very similar to that of the local Universe, although shifted to higher values of SFR. This indicates that, for the same stellar mass M$_{*}$, star forming galaxies at z$\sim$0.84 are under stronger star formation episodes than their local analogous.

The sSFR shows a negative correlation with stellar mass. The star formation in more massive galaxies, although with higher SFR, has less impact than in less massive ones, due to the large stellar mass already formed. The same trend is observed in the local Universe, though shifted to lower sSFRs. This is in good agreement with the {\em downsizing} scenario, in which massive galaxies are formed earlier than less massive ones. The fraction of massive galaxies (M$_{*}>$10$^{11}$M$_{\odot}$) undergoing strong star formation processes ({\em b}$>$2), $\sim$29\% at z$\sim$0.84 against $<$3\% at the local Universe, also supports this scenario. 

Finally, we have quantified the {\em downsizing} estimating the {\em quenching} mass at z$\sim$0.84 and at the local Universe based on the H$\alpha$ star formation rate.
We find that M$_{Q}$=1.0$^{+0.6}_{-0.4}\times$10$^{12}\, M_{\odot}$ (log~(M$_{Q}$/M$_{\odot}$)=12.0$\pm$0.2) at z$\sim$0.84 and M$_{Q}$=7.9$^{+1.9}_{-1.5}\times$10$^{10}\, M_{\odot}$ (log (M$_{Q}$/M$_{\odot}$)=10.9$\pm$0.1) in the local Universe. The evolution since the local Universe is out of doubt, with an increase in the {\em quenching} mass of an order of magnitude.

\acknowledgments

We thank Armando Gil de Paz for helpful discussions and comments.
We acknowledge support from the Spanish Programa Nacional de Astronom\'{i}a y Astrof\'{i}sica under grant AYA2009-10368. Partially funded by the Spanish MICINN under the Consolider-Ingenio 2010 Program grant CSD2006- 00070: First Science with the GTC. VV acknowledges support from the AstroMadrid Program CAM S2009/ESP-1496:``Astrof\'{i}sica y Desarrollos Tecnol\'{o}gicos en la Comunidad de Madrid'' funded by the Comunidad de Madrid and the European Union. PGP-G acknowledge support from the Ram\'{o}n y Cajal Program financed by the Spanish Government and the European Union. This work is based in part on observations made with the Spitzer Space Telescope, which is operated by the Jet Propulsion Laboratory, Caltech under NASA contract 1407. GALEX is a NASA Small Explorer launched in 2003 April. We gratefully acknowledge NASA’s support for construction, operation, and scientific analysis of the GALEX mission. This research has made use of the NASA/IPAC Extragalactic Database (NED) which is operated by the Jet Propulsion Laboratory, California Institute of Technology, under contract with the National Aeronautics and Space Administration. Based in part on data collected at Subaru Telescope and obtained from the SMOKA, which is operated by the Astronomy Data Center, National Astronomical Observatory of Japan. We acknowledge Edward L. Wright for his world wide web cosmology calculator \citep{Wright06}, which has been used during the preparation of this paper.

\bibliographystyle{apj}
\bibliography{referencias_paper}

\clearpage
\appendix

\section{Completeness level as a function of stellar mass}
\label{App:completeness}

The SFR completeness level for our sample depends on the stellar mass, due to the selection process, which is basically a selection in equivalent width. In \citetalias{Villar08} we simulated the selection process to determine the completeness level versus the line flux, but without considering the dependence on stellar masses, as it was unnecessary. The process consisted in the introduction of simulated galaxies in the real images to check wether or not they were recovered by the selection method. Fake galaxies were created with different line and continuum fluxes, sizes and inclinations in the real images. The outcome was the line flux completeness level. 

\begin{figure}[t]
\includegraphics[width=15cm]{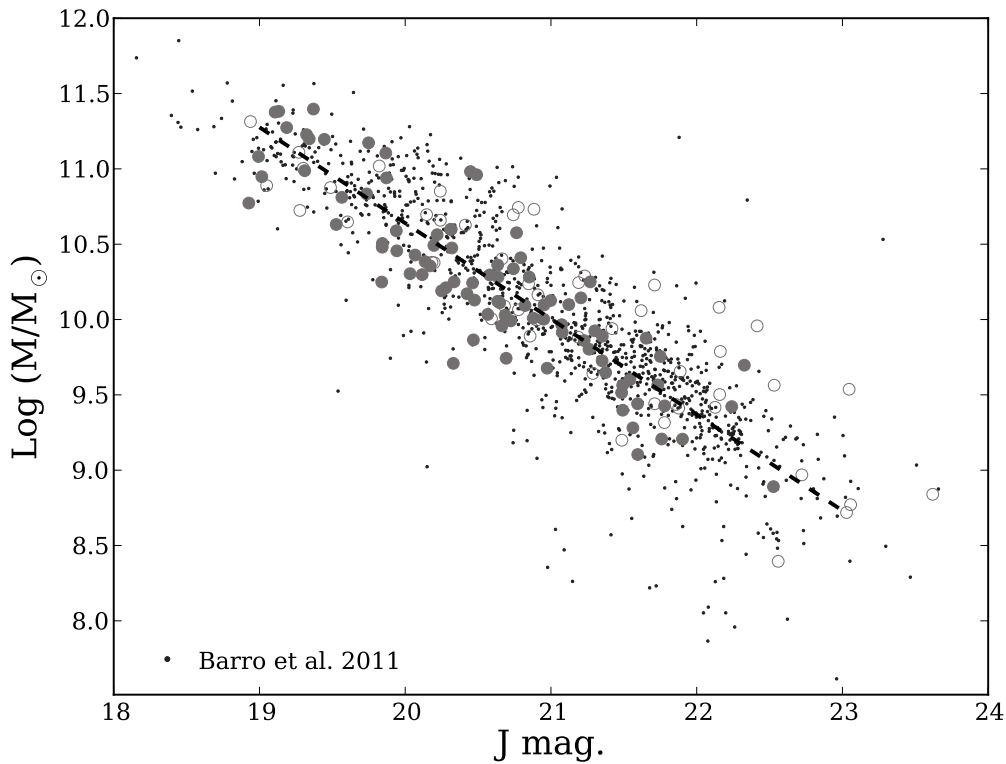}
\caption{\label{figure:j_mass} Stellar mass versus J band magnitude. The circles represent our sample: filled for the spectroscopic confirmations and open for the remainder. The small dots are the objects within our redshift range from the \cite{Barro11} mass selected sample. The dashed line is the best linear fit for the mass selected sample.   }
\end{figure}

The problem we need to address here is at which SFR, corrected for attenuation, the completeness level is 50\% as a function of stellar mass. The main problem is that we need to relate the stellar mass and extinction to the observables, i.e. the broad-band and narrow-band magnitudes. Intuitively, one may think that exists a correlation between the stellar mass and the broad band magnitude. Indeed, as can be seen in figure~\ref{figure:j_mass}, there is a correlation between the stellar mass and the J magnitude for our sample. However, it might be affected by selection biases, so we also check this relation with another sample. For this double-check we use the mass selected sample from the AEGIS database presented in \cite{Barro11} and \cite{Barro11b}. If we take all the objects with spectroscopic redshift within the limits of our sample we obtain a very similar relation. The best fit to the data gives: M$^{*}$=23.36\, -\, 0.636\, m$_{J}$.

With this correlation we can assign a stellar mass for an object given its J band flux. Thus, we can check the fraction of objects selected for a certain line-flux, defined by their emission in the broad and narrow band filters, and a stellar mass, estimated from the broad band flux. The line flux is transformed to SFR applying the corresponding calibration (equation~\ref{SFR_Halpha:eq}) and applying a mean correction for the nitrogen contribution ($I([NII]\lambda6584)/I(H\alpha)$=0.26). At the end we have a completeness level for each combination of SFR and stellar mass. From this we can obtain the SFR 50\% completeness level as a function of mass.

However, this completeness does not take into account the attenuation affecting the SFR measurement. To correct this effect we can apply the extinction correction to the SFR completeness level:

\begin{eqnarray}
SFR(M^{*})_{50\%}^{corrected} &=& SFR(M^{*})_{50\%}^{uncorrected} \ 10^{0.4\,A(H\alpha)}
\end{eqnarray}

\noindent where $SFR(M^{*})_{50\%}^{uncorrected}$ is the H$\alpha$ SFR for which we are 50\% complete at stellar mass M$^{*}$.

The question that arises now is which amount of extinction to apply. The immediate solution is to apply the mean(median) extinction correction obtained for the sample. However, it is well known that the amount of extinction depends on the total SFR. This fact has been shown for the \cite{Sobral09} sample by \cite{Garn10}. In particular these authors find that A(H$\alpha$)= $0.73 + 0.44 \log SFR(IR)$. As we saw in section~\ref{ir}, we can assume that the IR derived SFR is equal to the extinction corrected H$\alpha$ SFR, and thus we can write:

\begin{eqnarray}
SFR(M^{*})_{50\%}^{corrected} &=& SFR(M^{*})_{50\%}^{uncorrected} \ 10^{0.4\,(a+b\times \log SFR(M^{*})_{50\%}^{corrected})} \\ 
&=& SFR(M^{*})_{50\%}^{uncorrected} \ 10^{a'+b'\times \log SFR(M^{*})_{50\%}^{corrected}}
\end{eqnarray}

\noindent with a=0.73, b=0.44, a$'$=0.4$\times$a and b$'$=0.4$\times$b. We can find the value of SFR obtaining:

\begin{eqnarray}
\log SFR(M^{*})_{50\%}^{corrected} &=& \frac{\log SFR(M^{*})_{50\%}^{uncorrected} + a'}{1 - b'} 
\end{eqnarray}

This equation gives us the completeness level at 50\% considering that the extinction depends on the total SFR. The computed completeness curve is shown in figure~\ref{MASS_SFR}.

\end{document}